\documentclass[10pt,prb,aps,showpacs,twocolumn,unsortedaddress]{revtex4-1}
\usepackage[utf8]{inputenc}
\usepackage[english]{babel}
\usepackage{amsmath,amsfonts,amsthm,amsbsy, amssymb,bm}
\usepackage{mathtools}
\usepackage[caption=false]{subfig}
\usepackage{graphicx,bm}
\usepackage{physics}
\usepackage{dsfont}
\usepackage{xcolor}
\usepackage{adjustbox}
\usepackage{multirow}
\usepackage{breqn}
\usepackage{makecell}

\colorlet{linkequation}{red}
\usepackage[colorlinks,linkcolor=red,citecolor={blue},urlcolor={red}]{hyperref}

\makeatletter
\let\cat@comma@active\@empty
\makeatother
\begin{document}
	
	\title{Topological chiral interface states beyond insulators}

	\author{Lavi K. Upreti}
	\author{Pierre Delplace}

	\affiliation{Univ Lyon, ENS de Lyon, Univ Claude Bernard, CNRS, Laboratoire de Physique, F-69342 Lyon, France}

	\date{\today}
	\begin{abstract}
		%\textcolor{black}{We show how to engineer a chiral spectral flow of interface states in a gapless semi-metallic regime. Despite the bulk bands touch that makes the topological indices of the bands ill-defined, however, this non-trivial chiral spectral flow has a topological origin that is robust even when it coexists with the bulk bands. This phenomenon is illustrated with an adiabatically modulated 1D photonic quantum walk.}
			We propose a mechanism to engineer a chiral spectral flow of interface states in a gapless semi-metallic system. Although the topological indices of the bands  (Chern numbers) remain ill-defined, owing to the existence of bulk band touching points, this spectral flow has a topological origin that makes it robust even when it coexists with the bulk bands. These interface states between two gapless Dirac semimetals generalize the usual ones appearing  between two topologically distinct insulators. This phenomenon is illustrated with an adiabatically modulated photonic quantum walk
%. 
%		
%			
			
	\end{abstract}
	
	\maketitle

	\section{Introduction}

	Topological phases of matter involve both insulators \cite{Hasan2010} and semimetals \cite{ArmitageRMP}. However, their topological nature is revealed through different properties$ - $ 
while insulators are classified according to the global topology of their spectrally isolated bands over the Brillouin zone, semimetals are locally characterized by the topology of their nodal points or lines, that are somehow analogous to topological defects in reciprocal space. Interestingly, these two aspects are related. For instance, a Weyl semimetal can be seen as a stable phase when continuously deforming a 3D $\mathbb{Z}_2$ topological insulator into a trivial one, provided inversion symmetry is broken.\cite{Murakami2007,Murakami2007a} Also, in two dimensions, single Dirac points symptomatically appear at the transition between two distinct topological insulating phases, as in the celebrated Haldane model for Chern insulators.\cite{Haldane1988}

%This is the case of pair of Fermi arc surface states connecting the Weyl points when crossed by the Fermi energy in 3D.
The existence of boundary states is a universal manifestation of the underlying topology. For insulators, they are well defined in a spectral gap, that they bridge when varying their quasimomentum, which make them propagative. This property is sometimes referred to as the  \textit{spectral flow} of the boundary modes. It gives rise, for instance, to chiral edge states in Chern insulators. The absence of a spectral gap makes the situation quite different for semimetals. Indeed, the nodal points act as topological transition points in momentum space, at which the boundary states are therefore doomed to arise and end. This is the case of Fermi arcs surface states connecting the Weyl points when crossed by the Fermi energy in 3D. \cite{Burkov2011, Wan2011, Delplace2012, Xu2015} This is again the case of edge states in 2D graphene-like structures that own Dirac points in their energy spectrum.\cite{Fujita1996,Nakada1996}  
Obviously, the stability of the these Dirac points prevent the edge states from being chiral, so that the only way to get a spectral flow in 2D structures is to impose a gap, even  an indirect one,\cite{Palumbo2015,Jiangbin2016, Kamenev2018} as it is indeed the case for Chern insulators. Moreover, the Chern numbers $C$ of the bands of an insulator consistently account for the number of chiral edge states in each gap,  according to the celebrated bulk-edge correspondence.\cite{Hatsugai1993,Graf2013}

In this paper, we present a mechanism to engineer a spectral flow of interface states in 2D semimetallic phases with stable band touching (Dirac) points. In particular, we show how such interface states can spectrally coexist with bulk modes and still have a well defined topological origin although the Chern numbers of the bands $C$ are ill-defined.
 Such an unusual property is actually made possible by managing different gap inversions mechanisms at the same energy, so that all the gaps never open simultaneously. Even though there is no gap in the Brillouin zone, the topological properties can be captured by associating a topological charge as the quantized Berry flux emanating from degeneracy points in 3D parameter space, similarly to Weyl nodes in synthetic dimensions. This quantity is also a Chern number $\mathcal{C}$, but must be distinguished from that  of the full bands over the Brillouin zone $C$ which are ill-defined because the bands touch. Remarkably, the resulting topological spectral flows consist in chiral interface states in between 2D semimetals rather than insulators. All along the paper, we illustrate this idea with the quasienergy bands of a scattering network model that generalizes different experimental setups recently used in the context of topological photonic quantum walks.
\cite{Kitagawa2012, Bellec2017, Wimmer2017} In that context, the  parameter that tunes the spectral flow is an adiabatic phase driving the quantum walk, rather than the usual quasimomentum.

	The paper is organized as follows. In Section \ref{sec:scattering}, we introduce the scattering network model of an adiabatically modulated 1D quantum walk. That shows how the short time modulation of a dynamical phase allows one to selectively manipulate Dirac points at the same quasienergy. The topological properties of these degeneracy points are given in section \ref{sec:topo}, and their associated spectral flows of uncoupled interface states with bulk modes are computed accordingly. These \textit{interface} states are found to have no \textit{edge} state counterpart in finite geometry with open boundary conditions as discussed in section \ref{sec:edge}.

	\section{Scattering network model}
	\label{sec:scattering}
	
	\subsection{Generalities}
\begin{figure}[h!]
		\centering
		\includegraphics[width=0.5\textwidth]{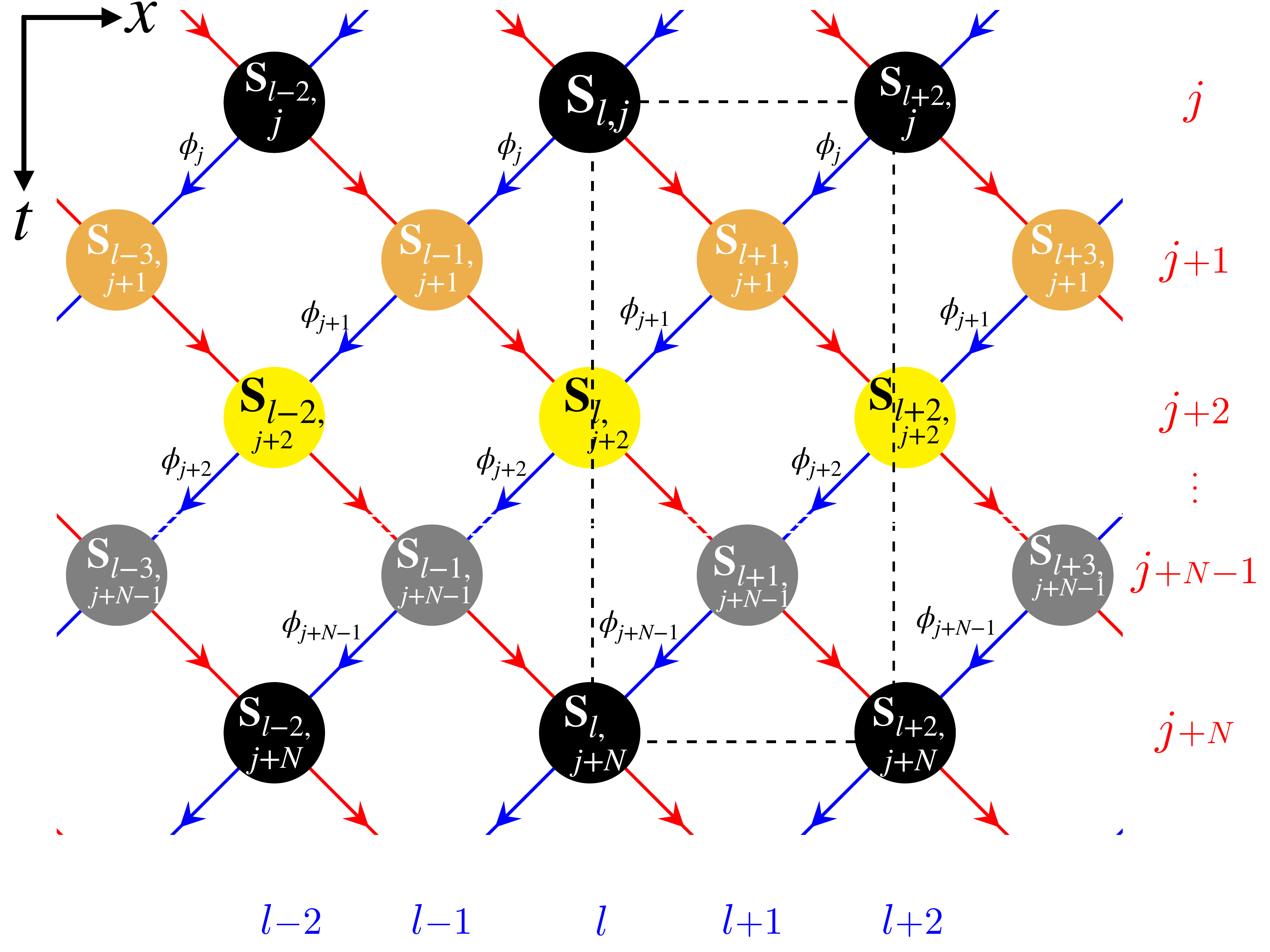}
		\caption{Two-dimensional oriented scattering lattice where the $y$ axis plays the role of time. A time period consists in $N$ successive steps. A phase $\phi_{j}$ is added for the states scattered out the node $j$ and propagating leftwards (blue arrows). The unit cell of this lattice is emphasized by a black dashed rectangle.}
		\label{fig:sc4}
	\end{figure}

Topological properties of networks models have been investigated in details recently\cite{Chong2013, Chong2014, Chong2015X, Tauber2015, Chong2016, Delplace2017, Delplace2019, Upreti2019, Potter2020} in particular in the context of photonics.
Here we consider the discrete time-evolution of a state (e.g. quantum state or wavepacket) through the generic network depicted in Fig.~\ref{fig:sc4}. Such evolution decomposes into a staggered sequence of free propagations along the oriented links and scattering events at the vertices encoded into the unitary matrices
\begin{eqnarray}\label{eq:sj4a}
	S_{l,j} &=& 
	\begin{pmatrix}
	\cos\theta_{l,j} & i\sin\theta_{l,j} \\ 
	i\sin\theta_{l,j}& \cos\theta_{l,j}
	\end{pmatrix}
	\end{eqnarray}
where the parameters $\theta_{l,j}$ may depend both on the position $l$ in the lattice and on the time step $j$. Let us impose a periodicity of the network after $N$ time steps so that the scattering nodes satisfy $S_{l,j}=S_{l,j+N}$. Such network can thus model e.g. a 1D Floquet quantum walk. In the following, it will be indeed useful to introduce the Floquet operator $U_F$, that is the unitary evolution operator after $N$ time steps. This operator will depend on the set of parameters $\{ \theta_j\}$ where $j$ runs from $1$ to $N$ and where we have dropped the position index $l$, assuming invariance along the $x$ direction for now. \textcolor{black}{This position dependence will be reintroduced in sections \ref{sec:flot1} and \ref{sec:flot2} when discussing the interfacial spectral flows.}

	In addition, we introduce a phase shift  carried along by the states when propagating along each link. Importantly, this phase shift evolves according to two time scales with respect to the period of $N$ time steps. 
	At short time scales, the phase shift $\phi$ may also depend on both the discrete position $l$ and the time step $j$ within a period. \textcolor{black}{For simplicity sake, we omit the position index from these phases, like $ \theta $'s.} Then each unit cell of the network is thus decorated with a pattern of phase shifts $\phi_{j}=Q_{j}\phi$ that preserves the periodicity of the network, where $Q_{j}$ is some rational number that will be specified later, and $\phi$ is a phase shift of reference. Moreover, the second time scale appears through the small variation of this $\phi$ from one period to the next one. Strictly speaking, the network is therefore not periodic in time anymore, but we shall consider the case where  $\phi$ evolves slowly enough so that  the long time stroboscopic dynamics can be described by the adiabatically modulated Floquet operator when continuously varying the phase parameter $\phi\in[0,2\pi]$. Exploiting translation invariance along the $x$ direction,  we are finally interested in the Bloch-Floquet parametrized evolution operator $U_F( k_x, \phi, \{ \theta_j\})$ where $k_x$ is the quasi-momentum in the $x$ direction. Note that this operator depends in a periodic fashion on its parameters. Its eigenstates are thus parametrized over a $N+2$-torus that can be seen as a synthetic Brillouin zone. This model could thus provide an interesting framework to investigate physical phenomena in higher dimensions ($>3$), such as analogs of topological phases provided a spectral gap exists.

	This network generalizes previous models whose topological properties have been investigated experimentally in photonics setups. For instance,  when $N=2$ and in the absence of a phase shift ($\phi=0$), the model describes one-dimensional photonic quantum walk \cite{Kitagawa2012} and one-dimensional laser-written modulated photonic lattices in silica,\cite{Bellec2017} in which boundary modes have been observed. Still when $N=2$ but now for  $ \phi_1 = + \phi, \phi_2 = - \phi $ together with fixed coupling parameters $ \theta_{l,j} = \pi/4 $, it describes   pairs of coupled optical fibre loops in which the Berry curvature was measured using wavepacket dynamics. \cite{Wimmer2017}
		
%	 if one considers just two steps in Fig~\ref{fig:sc4} with zero phase ($ \phi_{j=1,2} = 0 $), it gives rise to standard periodically time dependent SSH model, implemented in experiments, for example in quantum walks \cite{Kitagawa2012} or photonics waveguide arrays \cite{Bellec2017}. Moreover, quite recently periodically time modulated phi's as $ \phi_1 = + \phi, \phi_2 = - \phi $ along with fixed coupling parameters ($ \theta_{i} = \pi/4 $, known as beam splitter condition) were implemented in circular fibers \cite{Wimmer2017}, where non-trivial topological signature for example Berry curvature was mapped using wavepacket dynamics. These two things all together show the promising nature of formulating dynamics in scattering network platform and closeness to experimental setups. 
	
	Interestingly, as we show below, this model can also exhibit spectral flows in the absence of a spectral gap. A particular striking case even consists of the uncoupled superposition of a spectral flow with the bulk bands.  
	 The topological nature of this spectral flow can be understood from the existence of topological charges (Berry monopoles) in synthetic dimensions $(k_x,\phi, \Delta)$ where  $\Delta$ depends on the scattering parameters $\theta_j$. The key feature that yields the coexistence of bulk and boundary modes is the selective manipulation of distinct gaps opening/closure mechanisms that are made possible by specific patterns of $\phi_{j}$. 
	 %It follows for instance that the creation of chiral edge states can be tuned while keeping a specific gap closed, before, at and after the transition.

	\subsection{Gapless states in four-steps networks}
	\label{sec:gapless}

\begin{figure}[htb!]
			\includegraphics[scale=0.55]{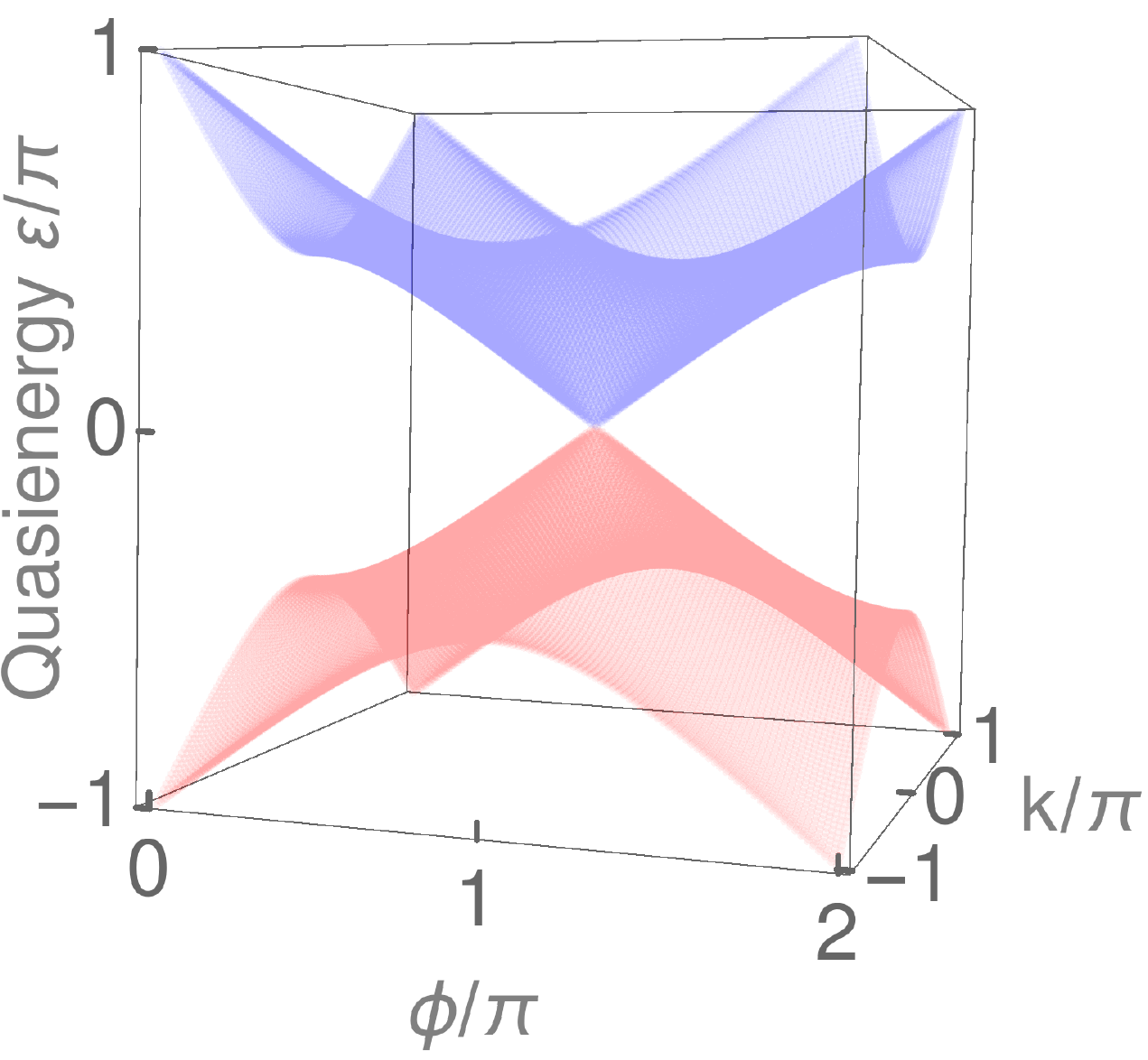}
			\caption{Quasienergy spectrum of the two-step Floquet operator $\phi_1 =-\phi_2 =  \phi $ and $ \theta_{j,l}=\pi/4 $, as considered in Ref\cite{Wimmer2017}}
			\label{fig:wim}
		\end{figure}

Our starting point is the two-step network ($N=2$) for which $\phi_1 =-\phi_2 =  \phi $ and where the nodes act as identical $50/50$ beam splitters.\cite{Wimmer2017} The quasi-energy spectrum of the Bloch-Floquet operator $U_F(k_x,\phi,\{\theta_i=\pi/4\})$, shown in figure~\ref{fig:wim}, is fully gapless and its two bands touch linearly along $k_x$ and $\phi$ at both quasi-energies $0$ and $\pi$. Tuning the coupling parameters away from $\theta=\pi/4$ opens a gap at quasi-energies $0$ and $\pi$, and may lead to a topological regime characterized by gap-valued topological invariants $W\in \pi_3(U(N))$.\cite{Rudner2013, Delplace2017} Accordingly, in the presence of boundaries in the $x$ direction, the system exhibits a spectral flow of edge states that bridges each gap when $\phi$ is tuned.\cite{Upreti2019}
Here we show how to implement a similar spectral flow \textcolor{black}{while bands touch}. For that purpose, one needs additional Dirac points at the same quasi-energies, and whose stability depends on different parameters than the preexisting ones.  
\textcolor{black}{More band touching points can be obtained in the synthetic Brillouin zone $(\phi,k_x)$ by folding the quasienergy spectrum. This is achieved by} enlarging the unit cell when allowing for more distinct couplings along either the transverse or the propagative direction.
Here we choose the second option by considering a period of $N=4$ steps for the fast dynamics, and fixing the distinct phase shifts inside a unit cell as $ \phi_1 = + \phi, \phi_2 =-\phi, \phi_3 = + \phi$, and $\phi_4 = -\phi$. Note that the net phase inside the unit cell is zero, thus conserving  inversion symmetry along synthetic dimension $\phi$, which avoids any winding of the quasienergy bands.\cite{Jiangbin2016,Upreti2019,Upreti2020}

    Using the spatial periodicity along $x$, i.e. $\theta_{l,j} = \theta_j$ the Floquet-Bloch evolution operator can be written as the succession of translation-like operations and local scattering processes as
       \begin{align} \label{eq:floquet_def}
     U_{F}(\phi, k_{x};\{\theta_{j}\}) = T_-S_{4} T_+ S_{3} T_- S_{2} T_+ S_{1}
     \end{align}
    where 
    \begin{subequations}
         \begin{align}
        &T_\pm=\begin{pmatrix}
     e^{i (k_{x}\pm\phi)/2} & 0\\
     0 & e^{-i(k_{x}\pm\phi)/2}
     \end{pmatrix}\\
     &S_j=\begin{pmatrix}
    \cos\theta_{j} & i\sin\theta_{j} \\ 
    i\sin\theta_{j}& \cos\theta_{j}
    \end{pmatrix} \ .
     \end{align}  
     \end{subequations}

	\begin{figure}[htb!]
		\includegraphics[scale=0.42]{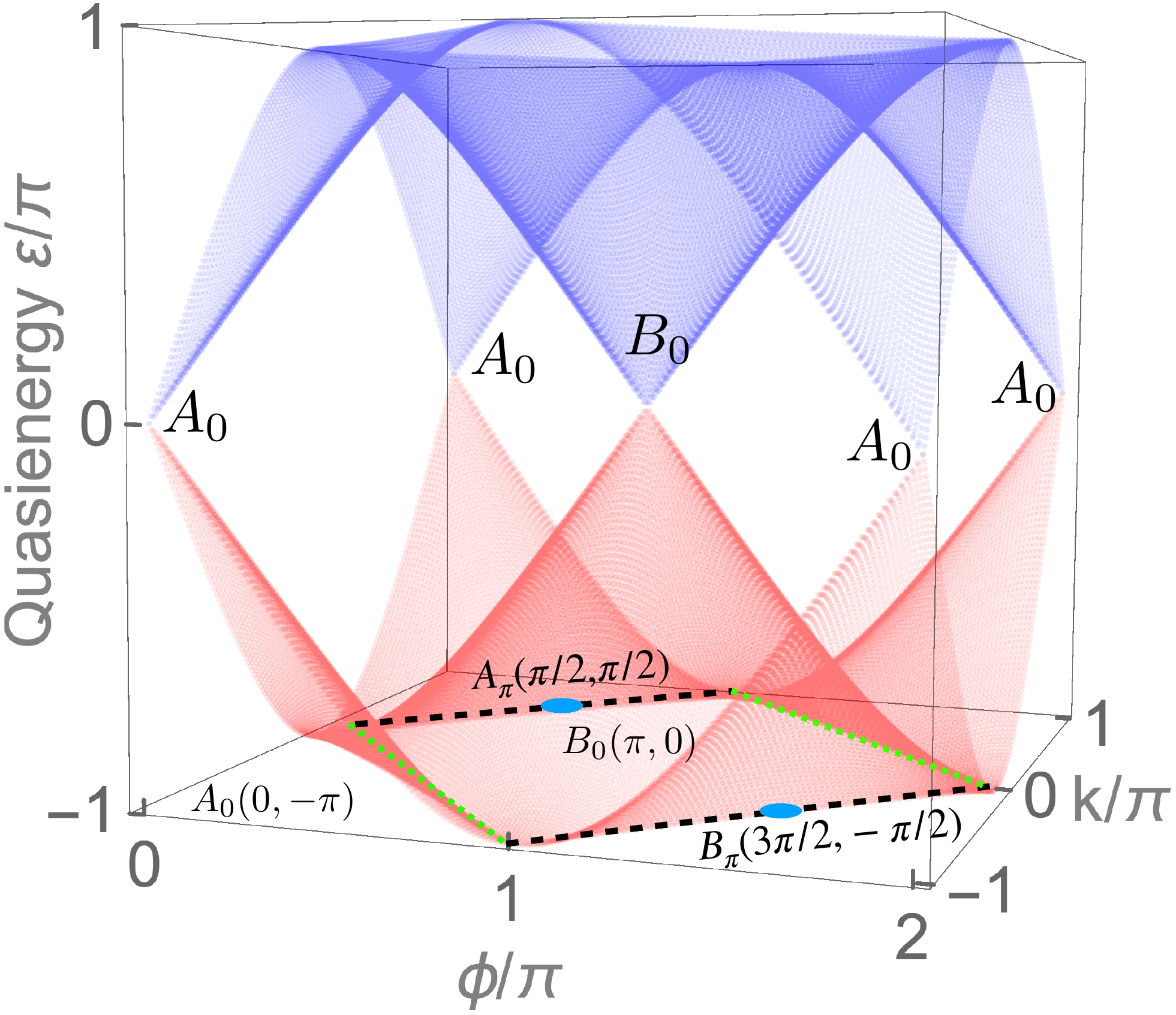}
		\caption{Quasienergy spectrum of the four-step Floquet operator  for $\phi_1 =-\phi_2 =\phi_3=-\phi_4  =\phi $ and  $ \theta_{j=1..4} = \pi/4 $. Bands touch at different points  at  $\varepsilon = 0$ and  along lines at $\varepsilon = \pi$. } 
		\label{fig:4dpi1}
	\end{figure}

As shown in Fig.~\ref{fig:4dpi1}, the quasienergy spectrum $\varepsilon$ is fully gapless for the critical value of parameters  $\{\theta_j=\pi/4\}$,  as expected. But now, there exists two Dirac points $ A_{0}, B_{0} $ at $\varepsilon=0$, and the two bands also touch at $\varepsilon=\pi$ along four lines (instead of points) satisfying $ k_{x}\pm\phi = 0 $ and  $ k_{x}\pm\phi = 2\pi$  (dashed black and green lines in Fig~\ref{fig:4dpi1}). By deviating from this critical point, it is then possible to lift some of these degeneracies leaving untouched the other ones. \textcolor{black}{The different  conditions to lift the specific band touching points (or lines) are inferred by expanding the Floquet operator in $\theta_j$ around these degeneracies.}

At the two Dirac points sitting at  $\varepsilon = 0 $, namely $A_0$ and $B_0$,  the Floquet operator must satisfy $U_F=\mathds{1}$. Substituting their coordinates $(\phi,k_x)$ --  respectively  $(0, \pi)$ and $(\pi,0)$ --  in equation\eqref{eq:floquet_def}, an expansion in scattering parameters around the critical point $\{\theta_j=\pi/4 \rightarrow \pi/4 +\delta \theta_j\}$, yields the constraint
 \begin{align} \label{eq:floquet_approx1}
   S(-\delta\theta_{1}+\delta\theta_{2}-\delta\theta_{3}+\delta\theta_{4}) =\mathds{1}  \quad \text{at}\ \, A_0\ \,\text{and}\,\ B_0
 \end{align}
(where $S(\theta_j)\equiv S_j$), which is only satisfied for $-\delta\theta_{1}+\delta\theta_{2}-\delta\theta_{3}+\delta\theta_{4}=0$. Conversely, a gap opens at $A_0$ and $B_0$ when this condition is not fulfilled.
An interesting twist comes at $\varepsilon = \pi $, where now the Floquet operator must satisfy $ U_{F}=-\mathds{1}$. Expanding the Floquet operator in scattering parameters for the four degeneracy lines $ k_{x}\pm\phi=0 $ and  $ k_{x}\pm\phi = 2\pi$ yields the condition
 \begin{align} \label{eq:floquet_approx2}
 S(\delta\theta_{1}+\delta\theta_{2}+\delta\theta_{3}+\delta\theta_{4}) = \mathds{1} 
 \end{align}
that differs from the condition  \eqref{eq:floquet_approx1} for the two degeneracy points at  $\varepsilon = 0 $. Furthermore, there are two special points, namely $ A_{\pi}$ at $(\phi = \pi/2,\ k_{x} = \pi/2) $ and $ B_{\pi}$ at $(\phi = 3\pi/2,\ k_{x} = -\pi/2) $, shown with blue dots in Fig~\ref{fig:4dpi1}, where this expansion does not apply. There, one finds a third condition that reads
 \begin{align} \label{eq:floquet_approx3}
 S(\delta\theta_{1}-\delta\theta_{2}-\delta\theta_{3}+\delta\theta_{4}) = \mathds{1}  \quad \text{at}\ \, A_\pi\ \,\text{and}\,\ B_\pi \ .
 \end{align}
Finally, the different gap opening terms $\delta \theta_j$ follow from 
 \begin{align} \label{eq:condi_gap}
\nu_{1}\delta\theta_{1}+\nu_{2}\delta\theta_{2}+\nu_{3}\delta\theta_{3}+\nu_{4}\delta\theta_{4} \neq 0
 \end{align}
with $\nu_j=\pm 1$ as summarized in Table~\ref{tab:mass}.
Thus, doubling the time period of the network indeed brings new degeneracies, namely, $A_{0,\pi}$ and $B_{\pi}$, \textcolor{black}{and preserving the pre-existing one that we denote by $ B_{0} $}. However, degeneracies at a fixed quasienergy, $0$ or $\pi$, are (un)stable under the same perturbations $\delta \theta_j$.  The only exception being at $ \varepsilon = \pi $ where the degeneracy lines (in eq\eqref{eq:floquet_approx2}) and degeneracy points (in eq\eqref{eq:floquet_approx3}) do not share the same stability, and hence can be gapped separately. \textcolor{black}{We can get rid of these degeneracy lines to end up with degeneracy points by changing the phase shift pattern $\phi_j$ within a period. These degeneracy points can be furthermore manipulated individually at the same quasienergy.}

	 \begin{table}[h]
	 	\centering
	 	\begin{tabular}{|c|c|c|c|c|c|}
	 		\hline 
	 		Quasienergy  & Degeneracy points & $ \nu_{1} $ & $ \nu_{2} $ & $ \nu_{3} $ & $ \nu_{4} $  \\  
	 		\hline 
	 		\multirow{2}{*}{$ \varepsilon = 0 $}& $ B_{0} $ & $ - $ & $ + $ & $ - $ & $ + $  \\ 
	 		%\cline{2-7} 
	 		& $ A_{0} $ & $ - $ & $ + $ & $ - $ & $ + $   \\ 
	 		\hline
	 		\multirow{3}{*}{$ \varepsilon = \pi $}&\makecell{$ k_{x}\pm\phi = 0,2\pi $ \\ (excluding $ A_{\pi}, B_{\pi} $)}  & $ + $ & $ + $ & $ +$ & $ + $   \\ 
	 		%\cline{2-7} 
	 		& $ A_{\pi} $ & $ + $ & $ - $ & $ - $ & $ + $   \\ 
	 		%\cline{2-7} 
	 		& $ B_{\pi} $ & $ + $ & $ - $ & $ - $ & $ + $    \\ 
	 		\hline 	 	
	 	\end{tabular} 
	 	\caption{Stability of the different band touchings (points or lines) under a perturbation  $\nu_j \delta \theta_j$\textcolor{black}{, as defined in eq.\eqref{eq:condi_gap}, for $\phi_1 =-\phi_2 =\phi_3=-\phi_4  =\phi $.}}
	 	\label{tab:mass}
	 \end{table}

	\subsection{Selective manipulation of degeneracies instabilities}

\begin{figure}[htb!]
		\includegraphics[scale=0.65]{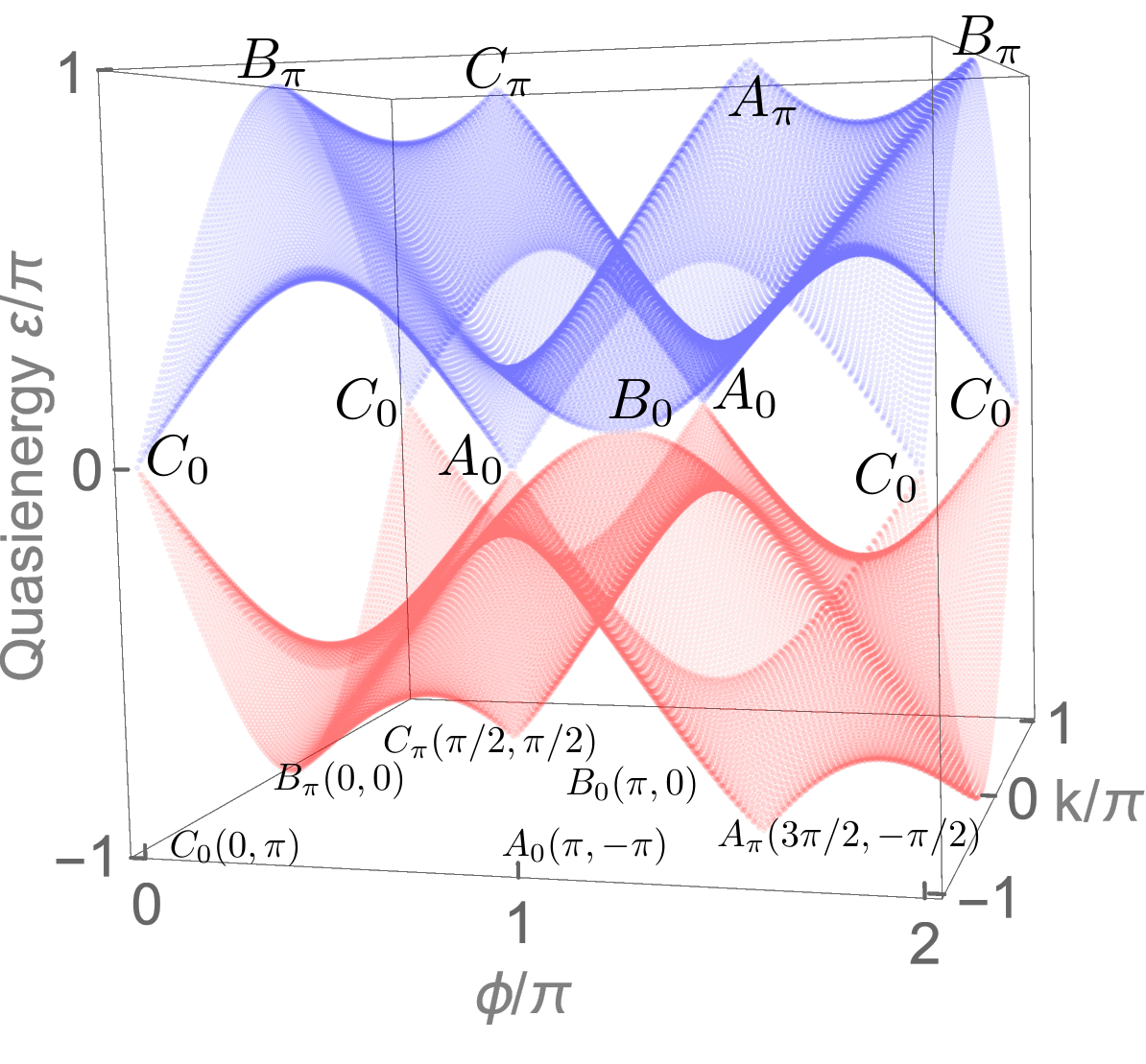}
		\caption{Quasienergy spectrum for the four-step Floquet operator with $ \phi_1 = 2 \phi, \phi_2 =-\phi, \phi_3 = 0$, and $\phi_4 = -\phi$ and $ \theta_{j=1..4} = \pi/4 $. It shows Dirac points at $ A_{0/\pi} $, $ C_{0/\pi} $ and semi-Dirac points at $ B_{0/\pi} $}		\label{fig:4dpi}
	\end{figure}

We propose now the following phase shift pattern that decorates the four-step period~: $ \phi_1 = 2 \phi, \phi_2 =-\phi, \phi_3 = 0$, and $\phi_4 = -\phi$. This choice still preserves $\sum \phi_j=0$ and thus prevents windings of the quasi-energy bands.\cite{Upreti2019,Upreti2020} The quasienergy spectrum of the Bloch-Floquet operator is depicted in Fig.~\ref{fig:4dpi} at the critical point $\{ \theta_j=\pi/4\}$. This spectrum is still fully gapless, but now the two bands touch at $\varepsilon=0$ and $\varepsilon=\pi$ only at points, either linearly in both directions (Dirac points $A_{0}, C_{0}, A_{\pi}$ and $C_{\pi}$) or linearly in one direction and quadratically in the other one  (semi-Dirac points $B_{0}$ and  $B_{\pi}$ \cite{Pickett,Gilles,Vanderbilt,Banerjee2015,Zhong2017,Mawrie2019}).

Applying the same reasoning as in  section \ref{sec:gapless} regarding the stability of these six band touching points with respect to a perturbation in scattering parameters $\delta \theta_j$, one ends up with a new classification shown in Table \ref{tab:mass2}.  It reveals four distinct gap opening processes. In particular, $C_0$, $A_\pi$ and $C_\pi$ behave similarly under any perturbation $\delta \theta_j$, but differently than the points $A_0$, $B_0$ and $B_\pi$. In other words, these different degeneracy points are  stable against distinct perturbations (or mass terms). It is thus now possible to lift a specific degeneracy at $\varepsilon=0$ or $\varepsilon=\pi$  without opening a bulk gap. For the sake of simplicity, instead of considering any combination of all the $ \delta\theta_{j} $'s, we fix $ \delta\theta_{1}=\delta\theta_{2} =0$, and focus only on the effect of $\delta \theta_3$ and $\delta \theta_4$. The stability of the degeneracies under the perturbations $\{\theta_3, \theta_4 \}\rightarrow \{\pi/4 + \nu_3 \delta \theta_3, \pi/4 + \nu_4\delta \theta_4\}$ can then be characterized by the sign of the product of $\nu_3 \nu_4$ only, that leaves us two possibilities. Therefore, one needs to distinguish two distinct gap opening processes driven by two independent mass terms 
\begin{align}\label{eq:mass_def}
m_\pm \equiv (\delta\theta_{3} \pm \delta\theta_{4})/2
\end{align}
 as summarized in Table ~\ref{tab:mass2} (in blue). These two mass terms allow us to generate topological spectral flows of boundary modes in a gapless (semi-metallic) regime, where both the Chern numbers and the Floquet winding number of the evolution operator are ill-defined, as we show in the next section.
	
	\begin{table}[h!]
		\centering
		\begin{tabular}{|c|c|c|c|c|c|c|}
			\hline 
			Quasienergy  & Degeneracy points & $ \nu_{1} $ & $ \nu_{2} $ & $ \textcolor{blue}{\nu_{3}} $ & $ \textcolor{blue}{\nu_{4}} $ & \textcolor{blue}{mass term}   \\ 
			\hline 
			\multirow{3}{*}{$ \varepsilon = 0 $}& $ A_{0} $ & $ - $ & $ - $ & $ \textcolor{blue}{-} $ & $ \textcolor{blue}{-} $ & $\textcolor{blue}{-m_+}$   \\ 
			\cline{2-7} 
			& $ B_{0} $ & $ + $ & $ - $ & $ \textcolor{blue}{-}  $ & $ \textcolor{blue}{+} $ & $\textcolor{blue}{-m_-}$  \\ 
			\cline{2-7} 
			& $ C_{0} $ & $ + $ & $ - $ & $ \textcolor{blue}{+} $ & $ \textcolor{blue}{-} $ & $\textcolor{blue}{m_-}$  \\ 
			\hline
			\multirow{3}{*}{$ \varepsilon = \pi $}& $ A_{\pi} $& $ + $ & $ - $ & $\textcolor{blue}{+} $ & $ \textcolor{blue}{-} $ & $\textcolor{blue}{m_-}$ \\ 
			\cline{2-7} 
			& $ B_{\pi} $ & $ + $ & $ + $ & $ \textcolor{blue}{-} $ & $ \textcolor{blue}{-} $ & \textcolor{blue}{$-m_+$} \\ 
			\cline{2-7} 
			& $ C_{\pi} $ & $ + $ & $ - $ & $ \textcolor{blue}{+} $ & $ \textcolor{blue}{-}$  & $\textcolor{blue}{m_-}$ \\ 
			\hline 	 	
		\end{tabular} 
		\caption{Stability of the different band touching points of Fig.~\ref{fig:4dpi}, under a perturbation $\nu_j \delta \theta_j$, \textcolor{black}{ as defined in Eq.\eqref{eq:condi_gap}, for $\phi_1/2 =-\phi_2 = -\phi_4  =\phi, \phi_3= 0 $}. The mass terms $m_\pm$ \textcolor{black}{ (from Eq.\eqref{eq:mass_def})} encode this stability when considering $\nu_3 \delta \theta_3$ and $\nu_4 \delta \theta_4$ only.}
		\label{tab:mass2}
	\end{table}

	%%%%%%%%%%%%%%%%%%%%%%%%%%%%%%%%%%%%%%%%%%%%%%%%%%%%
	\section{Topological spectral flow through bulk modes}
	\label{sec:topo}

	\subsection{Topological charge of degeneracy points}
	 \label{sec:top_charge}
	
In the vicinity of each band touching point $X$, one can expand the (dimensionless) effective Hamiltonian defined via the Floquet operator as
\begin{align}
U_F=e^{-iH_{\text{eff}}^{X}}
\end{align}
 at the lowest order terms in couplings $\delta \theta_j$, phase shift $\delta \phi$ and  quasimomentum $\delta k_{x}$. Such Hamiltonians have the generic form
 \begin{align}
 H^{X}_{\text{eff}} (\delta\phi, \delta k_{x}, m) = \mathbf{h}^{X}  \cdot \bm{\sigma}
 \end{align} 
where $\bm{\sigma}$ is the vector of Pauli matrices and \textcolor{black}{$\mathbf{h}^{X} = \{h_{1}^{X},h_{2}^{X},h_{3}^{X}\}$} defines a family of continuous maps from $\mathbb{R}^3$ to $\mathbb{R}^3$. Therefore, $\mathbf{h}^{X}/|\mathbf{h}^{X}|$ defines continuous maps from parameter space $ \mathbb{R}^{3}\backslash\{X\}$ to target space $S^2$ that are  classified by the homotopy group $\pi_2(S^2) = \mathbb{Z}$. The elements of this group are integer numbers that tell how many times  $\mathbf{h}^{X}/|\mathbf{h}^{X}|$ wraps the sphere. They are given by the degree of $\mathbf{h}^{X}$ defined as 
\begin{align}\label{eq:degree}
	\deg \mathbf{h}^{X} = \sum_{p_i^{(0)}} \text{sgn}\left[\det\left(\dfrac{\partial h_j^{X}}{\partial \lambda_{i}}\right)|_{h^{(0)}}\right]
	\end{align}
where the pre-images $p_i^{(0)}=(\delta\phi_{i}^{(0)}, \delta k_{xi}^{(0)}, m_{i}^{(0)})$ \textcolor{black}{of $\mathbf{h}^{(0)}$, an arbitrary vector in $\mathbb{R}^3$, satisfy by definition} $\mathbf{h}(p_i^{(0)})=\mathbf{h}^{(0)}$, and where $\{\lambda_{i}\}$ stands for $\{\delta\phi, \delta k_{x},m_\pm\} $. For a two band Hamiltonian, this degree is directly related to the Chern number $\mathcal{C}_\pm$ of the continuous family of normalized eigenstates $\psi_\pm(\delta\phi, \delta k_{x}, m)$ of $H^{X}_{\text{eff}} $ as
\begin{align}
\mathcal{C}_\pm = \mp \deg \mathbf{h}^{X} \ .
\label{eq:chern}
\end{align}
Importantly, a non vanishing value of $\mathcal{C}_n$ is known to guarantee the existence of a spectral flow towards bands $n$ when the mass term ($m_\pm$ here) is varied in space and changes sign.\cite{Nakahara2003, Volovik2009,  Delplace2017a, Faure2019, Perrot2019, Marciani2019} This spectral flow usually consists in a unidirectional mode, localized where the mass term vanishes, and whose (quasi-)energy bridges a spectral gap when a parameter (here $\phi$) is tuned.

In the following, we compute this topological index (via the degree formula \eqref{eq:degree}) for different band touching points (Dirac and semi-Dirac) and check numerically that their value correctly predicts a spectral flow, even in the absence of a gap.

	\subsubsection{Spectral flow induced by a spatial variation of $m_+$}
	\label{sec:flot1}
		
According to Table \ref{tab:mass2}, the degeneracy points $A_{0}$ and $B_\pi$ are both lifted when $m_+\neq 0$. One can thus assign them a topological charge (in the sense of section \ref{sec:top_charge}) by computing the degree of their respective expanded effective Hamiltonian, with the parameter (base) space being $(\delta \phi, \delta k_{x}, m_+)$. 

\begin{figure}[!htb]
		\centering
		\includegraphics[width=0.9\linewidth]{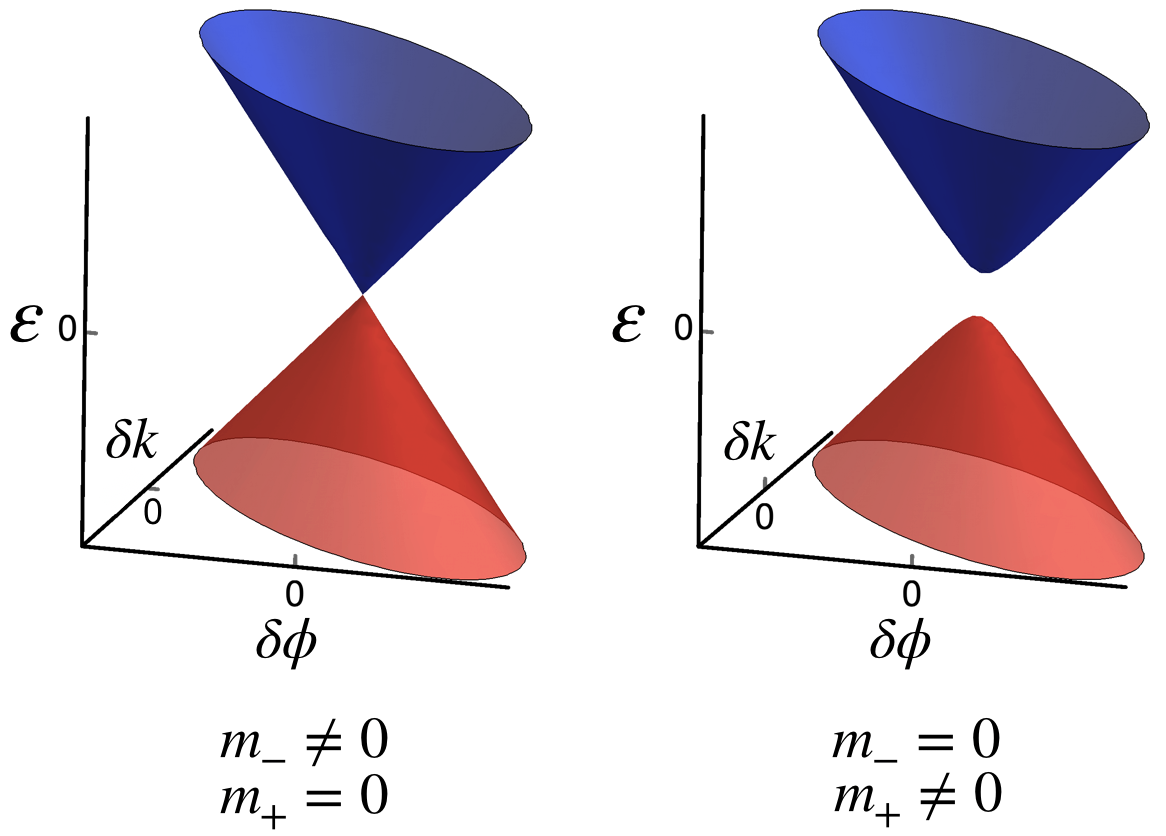}
		\caption{Dirac  points $A_0$ at quasienergy $ \varepsilon = 0 $ located at $ (\phi, k_{x}) = (\pi,\pi) $. This degeneracy point is robust against a perturbation $m_-$ but a gap opens due to the introduction of $m_+$.}
		\label{fig:diraca0}
	\end{figure}

Let us detail the calculation for $A_0$ whose coordinates  are $(\phi, k_x)=(\pi,\pi)\, [2\pi]$.
At lowest order in each parameter, the effective Hamiltonian $ H^{A_{0}}_{\text{eff}} = \mathbf{h}^{A_{0}} \cdot \bm{\sigma}$ yields
	\begin{align}\label{eq:hma0}
	\mathbf{h}^{A_{0}}(\delta\phi,\delta k_{x},m_+)=\
	\begin{pmatrix}
	-2\, m_+\\
	\delta \phi+\delta k_{x}\\
	\delta \phi \\
	\end{pmatrix} \ .
	\end{align}
	The spectrum of $ H^{A_{0}}_{\text{eff}} $ simply consists in the two branches $\varepsilon_\pm = \pm|\mathbf{h}^{A_{0}}|$  that touch linearly when $m_+=0$, as expected (see Fig \ref{fig:diraca0}). 	
Since $\mathbf{h}^{A_{0}}$ is linear with respect to each parameter, there is only one pre-image, so that the degree can be straightforwardly computed as
	\begin{align}\label{eq:hmass1}
	\deg \mathbf{h}^{A_{0}} =&\,  \text{sgn}\det\begin{pmatrix}
	\partial_{\delta\phi}h^{A_{0}}_{x}& \partial_{\delta\phi}h^{A_{0}}_{y} & \partial_{\delta\phi} h^{A_{0}}_{z}\\
	\partial_{\delta k_{x}}h^{A_{0}}_{x}&\partial_{\delta k_{x}}h^{A_{0}}_{y}  &  \partial_{\delta k_{x}}h^{A_{0}}_{z}\\
	\partial_{m_+}h^{A_{0}}_{x}& \partial_{m_+}h^{A_{0}}_{y} & \partial_{m_+}h^{A_{0}}_{z}
	\end{pmatrix}\nonumber\\
	=&\, \text{sgn}\det\begin{pmatrix}
	0 & 1 & 1 \\
	0 & 1 & 0 \\
	\textcolor{black}{ -2} & 0 & 0 \\
	\end{pmatrix}\notag \\
	=& +1
	\end{align}
Likewise, the topological charge for $ B_{\pi} $ is $ \deg \mathbf{h}^{B_{\pi}} = +1 $. 

\begin{figure}[!htb]
 		\centering
 		\includegraphics[width=1\linewidth]{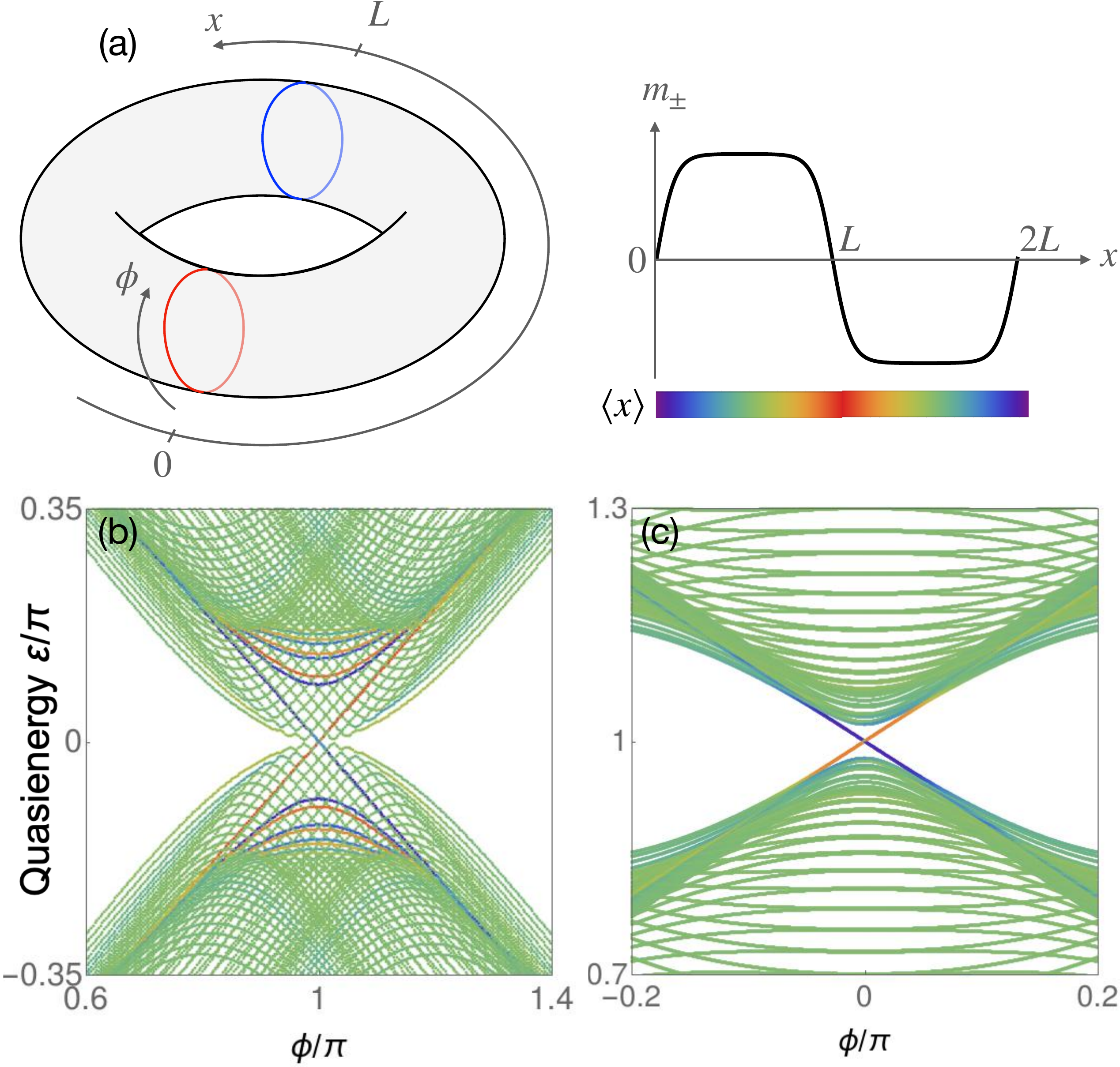}
 		\caption{Existence of chiral modes at the interface \textcolor{black}{(a) in a periodic geometry in $ x $ with 74 unitcells (from Fig.~\ref{fig:sc4}), and $ \phi $ (vertical), i.e., torus}, where the mass term $ m_+ $ changes sign twice \textcolor{black}{(namely at $ x = L $ and $ 2L $). The normalized position expectation value (to determine the localization of the states) is encoded in the color bar. For e.g., the localized interface states are shown in red (blue), when the mass term changes sign from $ + $ to $ - $ at $ x = L $ ($ - $ to $ + $ at $ x = 0, 2L $) and the states delocalized in the bulk are shown in green}. This for degeneracy point \textcolor{black}{for $ \theta_{j=1,2} = \pi/4 $} (b) $ A_{0} $ at quasienergy $0$, which exists along with the gapless $ B_{0} $ and another for (c) $ B_{\pi} $ at quasienergy $ \pi $.}	\label{fig:mplus}
 	\end{figure}

Accordingly, a spectral flow appears in the spectrum when considering now a spatial dependence of  $m_+(x)$ that changes sign. This anisotropy is taken into account in the network model by considering a variation of $\delta \theta_{l,3}=\delta \theta_{l,4}=m_+(l)$ along $x$, thus breaking translation invariance. For numerical convenience, we consider periodic boundary conditions along $x$, so that the mass term $m_+(l)$ changes sign twice, giving rise to \textcolor{black}{a cylindrical geometry with} two opposite spectral flows (instead of one)\textcolor{black}{, as shown in Fig.~\ref{fig:mplus}(a), where modes localized at these two interfaces of the cylindrical geometry are shown in blue and red}. This is a typical situation where chiral interface states bridge a spectral gap. Note that the situation is however different from what is currently encountered in topological insulating phases, since the two bands actually touch at $\varepsilon=\pi$ at two other points of the Brillouin zone, $A_\pi$ and $C_\pi$,  that are stable against the perturbation in $m_+$. Therefore, these chiral states  cannot be interpreted as the interface modes between two distinct gapped topological (e.g. Chern) insulators, but rather as interface modes  between two gapless semi-metals. The situation is maybe even more unusual with $A_0$, since its $\phi$ coordinate matches that of $B_0$ (see Fig.~\ref{fig:4dpi}) which remains stable under the perturbation in $m_+$, according to Table \ref{tab:mass2}. It follows that the spectral flow coexists with bulk modes and does not bridge a gap, as shown in Fig.~\ref{fig:mplus}(b). Notice that the direction of the spectral flow is the same for $B_\pi$ and $A_0$, in agreement with the common value of their topological charge.
%
%     \begin{figure}[h!]         
%     	\subfloat[\label{sfig:A0}]{%
%     		\includegraphics[width=.5\linewidth]{interface_beta1}
%     	}
%     \subfloat[\label{sfig:Bpi}]{%
%     	\includegraphics[width=.5\linewidth]{interface_beta_bpi}
%     }
%     	\caption{Existence of chiral modes at the interface, where the mass term $ m_+ $ changes sign twice with $x$ (periodic geometry). For degeneracy point (a) $ A_{0} $ at quasienergy $0$, which exists along with the gapless $ B_{0} $ and another for (b) $ B_{\pi} $ at quasienergy $ \pi $.}	\label{fig:mplus}
%     \end{figure}

		\subsubsection{Spectral flow induced by a spatial variation of $m_-$ }
		\label{sec:flot2}
				
\begin{figure}[!htb]
		\centering
		\includegraphics[width=0.9\linewidth]{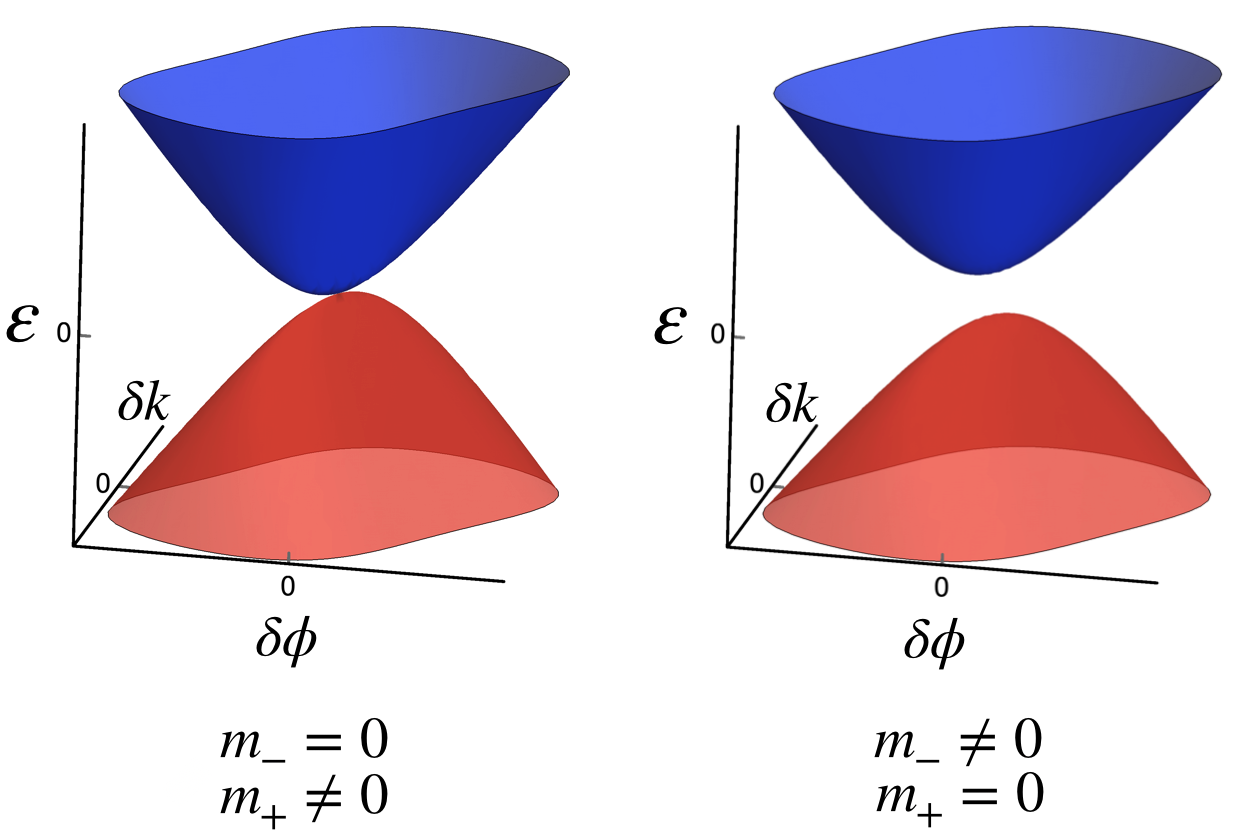}
		\caption{Semi-Dirac point $B_0$ at quasienergy $ \varepsilon = 0 $ located at $ (\phi, k_{x}) = (\pi,0) $. This degeneracy point is robust against a perturbation $m_+$ but a gap opens due to the introduction of $m_-$.}
		\label{fig:quad41}
	\end{figure}

Similarly, a spatial variation of the mass term $m_-$ leads to a topological spectral flow for $B_0$, $C_0$, $A_\pi$ and $C_\pi$ when $\phi$ is varied, provided $m_-$ changes sign. Let us focus on $B_0$ which is a  semi-Dirac point, since their topological charge and their associated spectral flow is overlooked in the literature in comparison to Dirac points.
An expansion of the effective Hamiltonian $ H^{B_{0}}_{\text{eff}} = \mathbf{h}^{B_{0}} \cdot \bm{\sigma} $ in coupling parameters and quasimomenta gives
	\begin{align}\label{eq:hqad1}
	\mathbf{h}^{B_{0}}(\delta\phi,\delta k_{x},m_-)=
	\begin{pmatrix}
	-2 m_--\frac{\delta k_{x}}{2} (\delta \phi+\delta k_{x})\\
	2 \delta \phi \, m_-   \\
	\delta \phi-\delta k_{x}\\
	\end{pmatrix} 
	\end{align}
	The  eigenvalues   $\varepsilon_\pm = \pm |\mathbf{h}^{B_{0}}|$ yields a semi-Dirac behavior when $m_-=0$, as announced (see Fig. \ref{fig:quad41}). The introduction of $m_-$ opens a gap, and allows us to define the topological charge of this degeneracy point as
	\begin{align}\label{eq:hmquad1}
		\deg \mathbf{h}^{B_{0}} =& \sum_{p_i^{(0)}} \text{sgn}\det\begin{pmatrix}
			-\frac{\delta k_{x}}{2} & 2m_- & 1 \\
			 -\delta k_{x}-\frac{\delta \phi}{2} & 0 & -1 \\
			-2 & 2\delta \phi  & 0 \\
		\end{pmatrix}\notag \\
		=& \sum_{p_i^{(0)}}\text{sgn}\left[4m_--(3\delta k_{x} +\delta\phi)\delta\phi \right]
	\end{align}

One can evaluate the pre-images by fixing a direction for $\mathbf{h}^{B_0}$, say along $z$. This imposes the three following conditions
\begin{subequations}
\begin{align}
-4m_-&= \delta k_{x}(\delta \phi +\delta k_{x}) \\
\delta \phi\, m_- &=0\\
\delta\phi &> \delta k_{x}
\end{align}
\end{subequations}
Three pre-images $(\phi_i, k_{x,i}, m_{i})$ are found to satisfy these conditions~: $p_1^{(0)}=(-k_1, k_1, 0)$ with $k_1>0$, $p_2^{(0)}=(0, k_2, -k_2^2/4)$ with $k_2<0$ and $p_3^{(0)}=(\phi_3, 0, 0)$ with $\phi_3>0$. The pre-image $p_1^{(0)}$ yields a positive contribution to the sum \eqref{eq:hmquad1} while both $p_2^{(0)}$ and $p_3^{(0)}$ contribute negatively, so that finally $\deg\mathbf{h}^{B_{0}}=-1$.
 A similar calculation leads to  $\deg \mathbf{h}^{C_{\pi}}= +1$, $\deg \mathbf{h}^{A_{\pi}}= +1$ and $\deg \mathbf{h}^{C_0} = +1$. 

\begin{figure}[!htb]
    	\centering
    	\includegraphics[width=1\linewidth]{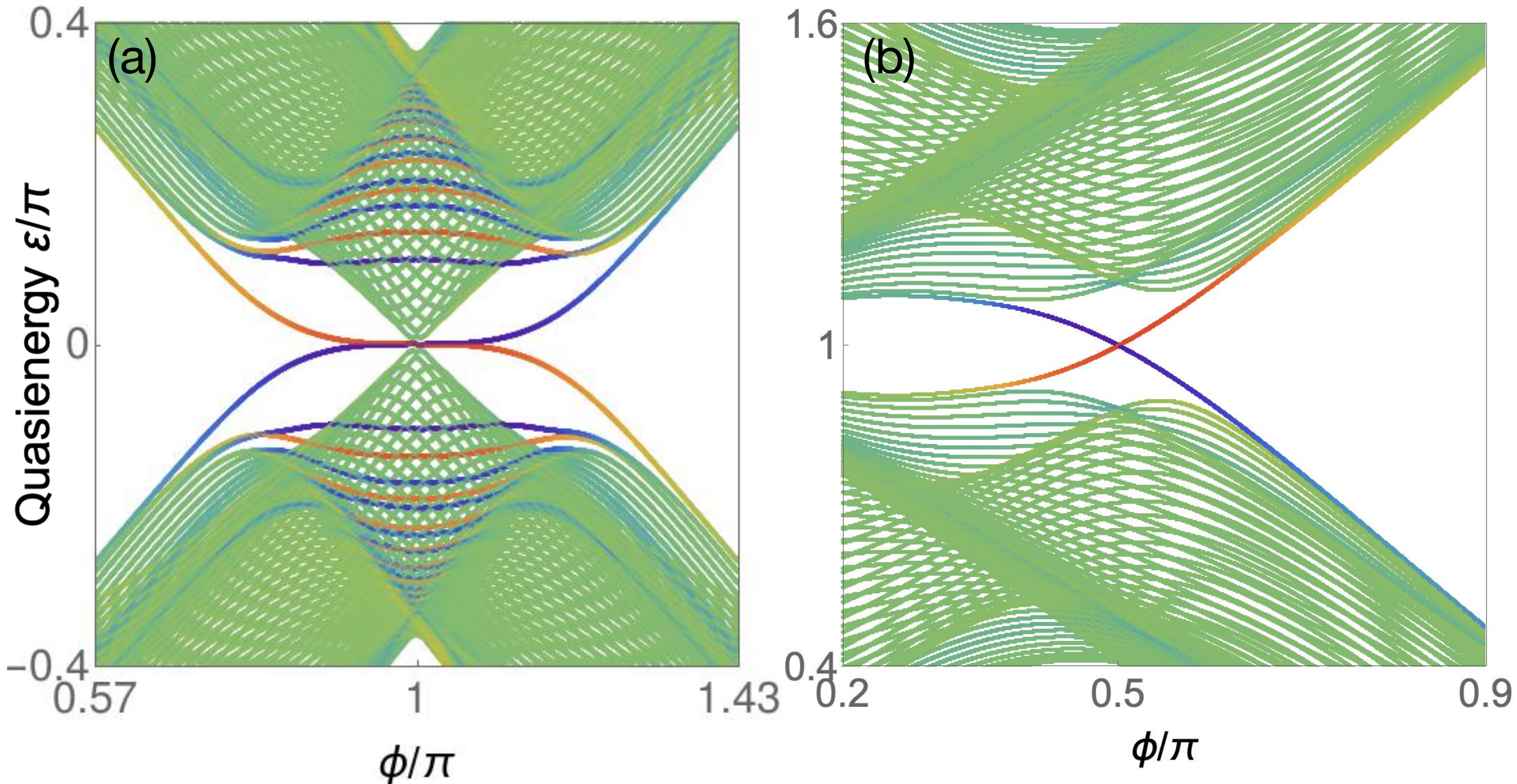}
    	\caption{Existence of chiral modes at the interface \textcolor{black}{for same periodic geometry (see Fig.~\ref{fig:mplus}(a)),} where the mass term $m_-$ changes sign twice, in the vicinity of (a) $ B_{0}$ at quasienergy $0$, and (b) $C_{\pi} $  at quasienergy $ \pi $. In the first case, the spectral flow crosses the Dirac point  $ A_{0} $. \textcolor{black}{The geometry and color code are that of Fig.~\ref{fig:mplus}}}\label{fig:mminus}  	
    \end{figure}

Accordingly, a numerical calculation is performed in a periodic geometry where $m_-(l)$ changes sign twice when varying with the discrete position index $l$ on the network. Spectral flows are found in agreement with the value of the topological charge. The cases of the semi-Dirac points $C_\pi$ and $B_0$ are depicted in \textcolor{black}{Fig.~\ref{fig:mminus}}, where the spectral flow for $B_0$ is indeed the opposite to that of $C_\pi$, \textcolor{black}{i.e., the slope of these modes in the quasienergy spectrum is opposite in sign for a given interface}. They both illustrate the same phenomenology as that discussed in the previous section in \textcolor{black}{Fig.~\ref{fig:mplus}}. In particular, there is a spectral flow around $B_0$ that coexists with bulk states, since $A_0$, that has the same $\phi$ coordinate,  is stable under a perturbation of $m_-$ type. Note the peculiar flattened dispersion relation of the interface states in Fig.~\ref{fig:mminus}  (a) which is reminiscent of the quadratic dispersion relation of the semi-Dirac point $B_0$ along the phi coordinate.

%     \begin{figure}[h!]
%     	\subfloat[\label{sfig:eta2}]{%
%     		\includegraphics[width=.535\linewidth]{interface_eta_cpi}
%     	}
%	  	\subfloat[\label{sfig:eta1}]{%
%     		\includegraphics[width=.5\linewidth]{interface_eta}
%     	}
%     	\caption{Existence of chiral modes at the interface, where the mass term $m_-$ changes sign twice, in the vicinity of (a) $C_{\pi} $  at quasienergy $ \pi $, and (b) $ B_{0}$ at quasienergy $0$. In the second case, the spectral flow crosses the Dirac point  $ A_{0} $. \textcolor{black}{The geometry and color code are that of Fig.~\ref{fig:mplus}}}\label{fig:mminus}  	
%     \end{figure}

 %%%%%%%%%%%%%%%%%%%%%%%%%%%%%%%%%%%%%%%%%%%%%%
 \section{Chiral edge states in gapless systems} 
 \label{sec:edge}

      \begin{figure}[htb!]
		\hspace*{-0.20cm}
		\includegraphics[scale=0.365]{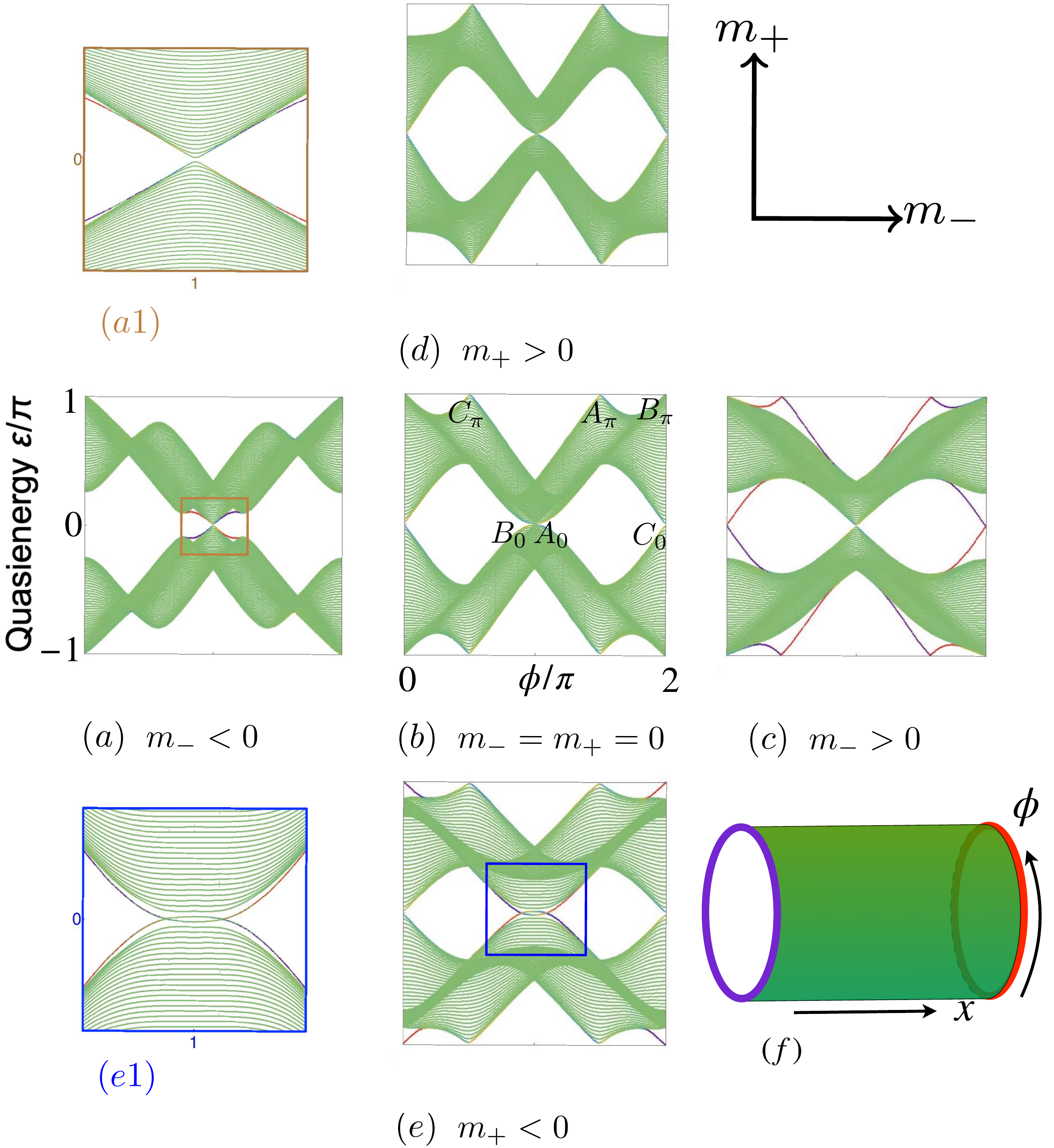}
		\caption{Quasienergy spectra of the four-step Floquet operator  for \textcolor{black}{$\phi_1/2 =-\phi_2 =-\phi_4  =\phi, \phi_3=0 $ and $\theta_{j=1,2} = \pi/4$} in finite geometry as a function of either $m_+$ or $m_-$. Imposing a vanishing of either $m_+$ or $m_-$ prevents a gap opening at $0$ and $\pi$ so that all the spectra are gapless. The two insets $ (a1) $ and $ (e1) $ show   edge states at $\varepsilon =0$ that merge to the bulk bands and disappear, unlike other chiral edge states for $m_\pm<0$ that live in local gaps.  \textcolor{black}{$ (f) $ Here, the finite geometry contains 50 unitcells while being periodic in $ \phi $, where the edge states localized at the left (right) boundary are in blue (red). This color coding is calculated from the normalized position expectation value (to determine the localization of the states), similar to in Fig.~\ref{fig:mplus}(a).} }
		\label{fig:4ftm}
	\end{figure}

\textcolor{black}{In topological insulators, there is commonly no fundamental distinction between edges and interfaces regarding the existence of boundary states: these are expected to automatically  exist in one case if they are shown to exist in the other. This is because an interface is seen as an edge between two insulators. The situation is different here with the semimetals we describe, since their topological property encoded through the index $\mathcal{C}$ is inherently related to an interface.}

The topological spectral flow described above consisting in confined modes at the \textit{interface} between two semimetallic regimes with mass terms  $m_\pm$ of   opposite signs, a natural question to ask is thus whether chiral \textit{edge} states may also exist at the boundary of a finite network with a fixed mass term $m_\pm$, as it is usually the case in topological insulators. However, in our configuration, the Chern number $C$ of the bands (to be distinguished from the Chern numbers assigned to the degeneracy $\mathcal{C}$ \eqref{eq:chern}) cannot be defined because of the band touching points in the bulk.

Quasienergy spectra computed in that geometry are shown in Fig.~\ref{fig:4ftm} for different values of a uniform $m_\pm$. We find chiral edge states (with respect to $\phi$) at $A_\pi$, $C_\pi$ and $C_0$ when $m_->0$, and at $B_\pi$ when  $m_+<0$.  Edge states are also found around $A_0$ for $m_+<0$ and $B_0$ for $m_-<0$, where the gap remains close.
The first remark is that  the sign of $m_\pm$ that gives rise to edge states does not seem obviously related to the topological charge computed above. Moreover, while these edge states look very similar to what can be found in gapped systems for $A_\pi$, $C_\pi$, $C_0$  and $B_\pi$, as they bridge a local gap, the situation is different for $A_0$ and $B_0$ that are affected by the bulk modes. The inset figures show that these edge states actually do not connect the two bands, but eventually couple to the bulk modes and disappear (Figs.~\ref{fig:4ftm}(a1) and (e1)). This is in sharp contrast with the continuous interfaces in $m_\pm$ that revealed a continuous spectral flow through the bulk modes (Figs.~\ref{fig:mplus}(b) and \ref{fig:mminus}(a)).

The existence of a spectral flow when a mass term is continuously varied and changes sign, is traditionally understood as a mode emerging at the interface between two topologically nonequivalent systems. This is of course meaningful provided that each system's topology is well defined in itself, when the mass term is fixed, like in Chern insulators. This is, however, not always the case. In particular, in continuous media, the Chern numbers $C$ of the bands  are only well defined when the projectors are regularized at infinity.\cite{Volovik1988,Silveirinha2015,Tauber2019,Souslov2019,Tauber2020} Otherwise, the topological charge approach used here remains a powerful valid strategy \cite{Delplace2017a,Perrot2019,Marciani2019}, but does not seem suitable to predict the edge state in a finite geometry as shown in figure \ref{fig:4ftm}.

	\section{Conclusion}

We have reported a model where the gapless spectrum prevents the definition of a Chern number for the bands and a Floquet winding number for the gaps.\cite{Rudner2013} Still, we have shown that topological chiral spectral flows of interface states can be engineered by a suitable anisotropic perturbation that changes sign. Such spectral flows can even coexist with delocalized (bulk) modes from which they remain uncoupled, in contrast to edge states in finite geometry with open boundaries. These spectral flows can be interpreted as robust chiral states at the interface between gapless semimetallic phases, \textcolor{black}{showing that a spectral gap is not necessary for chiral spectral flows to exist.}

\textcolor{black}{Pairs of coupled optical fibers \cite{Wimmer2013,AloisPRL2011,Miri2012,Regensburger2012,Wimmer2017} constitute an excellent candidate to engineer this new topological regime. Recently,  a spatial variation of coupling parameters ($ \theta $'s) has been achieved with an interface where the bulk gap closes \cite{Weidemann2020} in that setup. In addition, the temporal variation of the phase has also been achieved in the same setup \cite{Regensburger2012,Wimmer2017}. Both of these experimental results pave the way for our proposal. Coupled waveguide arrays, that demonstrated the versatility to manipulate original topological Floquet phenomena \cite{Szameit2010,Rechtsman2013,Bellec2017} might also be considered to implement this new regime.}

\section{Acknowledgements}

This work was supported by the French Agence Nationale de la Recherche (ANR) under grant Topo-Dyn (ANR-14-ACHN-0031).

	\bibliography{life}

\begin{thebibliography}{54}%
\makeatletter
\providecommand \@ifxundefined [1]{%
 \@ifx{#1\undefined}
}%
\providecommand \@ifnum [1]{%
 \ifnum #1\expandafter \@firstoftwo
 \else \expandafter \@secondoftwo
 \fi
}%
\providecommand \@ifx [1]{%
 \ifx #1\expandafter \@firstoftwo
 \else \expandafter \@secondoftwo
 \fi
}%
\providecommand \natexlab [1]{#1}%
\providecommand \enquote  [1]{``#1''}%
\providecommand \bibnamefont  [1]{#1}%
\providecommand \bibfnamefont [1]{#1}%
\providecommand \citenamefont [1]{#1}%
\providecommand \href@noop [0]{\@secondoftwo}%
\providecommand \href [0]{\begingroup \@sanitize@url \@href}%
\providecommand \@href[1]{\@@startlink{#1}\@@href}%
\providecommand \@@href[1]{\endgroup#1\@@endlink}%
\providecommand \@sanitize@url [0]{\catcode `\\12\catcode `\$12\catcode
  `\&12\catcode `\#12\catcode `\^12\catcode `\_12\catcode `\%12\relax}%
\providecommand \@@startlink[1]{}%
\providecommand \@@endlink[0]{}%
\providecommand \url  [0]{\begingroup\@sanitize@url \@url }%
\providecommand \@url [1]{\endgroup\@href {#1}{\urlprefix }}%
\providecommand \urlprefix  [0]{URL }%
\providecommand \Eprint [0]{\href }%
\providecommand \doibase [0]{http://dx.doi.org/}%
\providecommand \selectlanguage [0]{\@gobble}%
\providecommand \bibinfo  [0]{\@secondoftwo}%
\providecommand \bibfield  [0]{\@secondoftwo}%
\providecommand \translation [1]{[#1]}%
\providecommand \BibitemOpen [0]{}%
\providecommand \bibitemStop [0]{}%
\providecommand \bibitemNoStop [0]{.\EOS\space}%
\providecommand \EOS [0]{\spacefactor3000\relax}%
\providecommand \BibitemShut  [1]{\csname bibitem#1\endcsname}%
\let\auto@bib@innerbib\@empty
%</preamble>
\bibitem [{\citenamefont {Hasan}\ and\ \citenamefont {Kane}(2010)}]{Hasan2010}%
  \BibitemOpen
  \bibfield  {author} {\bibinfo {author} {\bibfnamefont {M.~Z.}\ \bibnamefont
  {Hasan}}\ and\ \bibinfo {author} {\bibfnamefont {C.~L.}\ \bibnamefont
  {Kane}},\ }\href {\doibase 10.1103/RevModPhys.82.3045} {\bibfield  {journal}
  {\bibinfo  {journal} {Rev. Mod. Phys.}\ }\textbf {\bibinfo {volume} {82}},\
  \bibinfo {pages} {3045} (\bibinfo {year} {2010})}\BibitemShut {NoStop}%
\bibitem [{\citenamefont {Armitage}\ \emph {et~al.}(2018)\citenamefont
  {Armitage}, \citenamefont {Mele},\ and\ \citenamefont
  {Vishwanath}}]{ArmitageRMP}%
  \BibitemOpen
  \bibfield  {author} {\bibinfo {author} {\bibfnamefont {N.~P.}\ \bibnamefont
  {Armitage}}, \bibinfo {author} {\bibfnamefont {E.~J.}\ \bibnamefont {Mele}},
  \ and\ \bibinfo {author} {\bibfnamefont {A.}~\bibnamefont {Vishwanath}},\
  }\href {\doibase 10.1103/RevModPhys.90.015001} {\bibfield  {journal}
  {\bibinfo  {journal} {Rev. Mod. Phys.}\ }\textbf {\bibinfo {volume} {90}},\
  \bibinfo {pages} {015001} (\bibinfo {year} {2018})}\BibitemShut {NoStop}%
\bibitem [{\citenamefont {Murakami}(2007)}]{Murakami2007}%
  \BibitemOpen
  \bibfield  {author} {\bibinfo {author} {\bibfnamefont {S.}~\bibnamefont
  {Murakami}},\ }\href {\doibase 10.1088/1367-2630/9/9/356} {\bibfield
  {journal} {\bibinfo  {journal} {New Journal of Physics}\ }\textbf {\bibinfo
  {volume} {9}},\ \bibinfo {pages} {356} (\bibinfo {year} {2007})}\BibitemShut
  {NoStop}%
\bibitem [{\citenamefont {Murakami}\ \emph {et~al.}(2007)\citenamefont
  {Murakami}, \citenamefont {Iso}, \citenamefont {Avishai}, \citenamefont
  {Onoda},\ and\ \citenamefont {Nagaosa}}]{Murakami2007a}%
  \BibitemOpen
  \bibfield  {author} {\bibinfo {author} {\bibfnamefont {S.}~\bibnamefont
  {Murakami}}, \bibinfo {author} {\bibfnamefont {S.}~\bibnamefont {Iso}},
  \bibinfo {author} {\bibfnamefont {Y.}~\bibnamefont {Avishai}}, \bibinfo
  {author} {\bibfnamefont {M.}~\bibnamefont {Onoda}}, \ and\ \bibinfo {author}
  {\bibfnamefont {N.}~\bibnamefont {Nagaosa}},\ }\href {\doibase
  10.1103/PhysRevB.76.205304} {\bibfield  {journal} {\bibinfo  {journal} {Phys.
  Rev. B}\ }\textbf {\bibinfo {volume} {76}},\ \bibinfo {pages} {205304}
  (\bibinfo {year} {2007})}\BibitemShut {NoStop}%
\bibitem [{\citenamefont {Haldane}(1988)}]{Haldane1988}%
  \BibitemOpen
  \bibfield  {author} {\bibinfo {author} {\bibfnamefont {F.~D.~M.}\
  \bibnamefont {Haldane}},\ }\href {\doibase 10.1103/PhysRevLett.61.2015}
  {\bibfield  {journal} {\bibinfo  {journal} {Phys. Rev. Lett.}\ }\textbf
  {\bibinfo {volume} {61}},\ \bibinfo {pages} {2015} (\bibinfo {year}
  {1988})}\BibitemShut {NoStop}%
\bibitem [{\citenamefont {Burkov}\ and\ \citenamefont
  {Balents}(2011)}]{Burkov2011}%
  \BibitemOpen
  \bibfield  {author} {\bibinfo {author} {\bibfnamefont {A.~A.}\ \bibnamefont
  {Burkov}}\ and\ \bibinfo {author} {\bibfnamefont {L.}~\bibnamefont
  {Balents}},\ }\href {\doibase 10.1103/PhysRevLett.107.127205} {\bibfield
  {journal} {\bibinfo  {journal} {Phys. Rev. Lett.}\ }\textbf {\bibinfo
  {volume} {107}},\ \bibinfo {pages} {127205} (\bibinfo {year}
  {2011})}\BibitemShut {NoStop}%
\bibitem [{\citenamefont {Wan}\ \emph {et~al.}(2011)\citenamefont {Wan},
  \citenamefont {Turner}, \citenamefont {Vishwanath},\ and\ \citenamefont
  {Savrasov}}]{Wan2011}%
  \BibitemOpen
  \bibfield  {author} {\bibinfo {author} {\bibfnamefont {X.}~\bibnamefont
  {Wan}}, \bibinfo {author} {\bibfnamefont {A.~M.}\ \bibnamefont {Turner}},
  \bibinfo {author} {\bibfnamefont {A.}~\bibnamefont {Vishwanath}}, \ and\
  \bibinfo {author} {\bibfnamefont {S.~Y.}\ \bibnamefont {Savrasov}},\ }\href
  {\doibase 10.1103/PhysRevB.83.205101} {\bibfield  {journal} {\bibinfo
  {journal} {Phys. Rev. B}\ }\textbf {\bibinfo {volume} {83}},\ \bibinfo
  {pages} {205101} (\bibinfo {year} {2011})}\BibitemShut {NoStop}%
\bibitem [{\citenamefont {Delplace}\ \emph {et~al.}(2012)\citenamefont
  {Delplace}, \citenamefont {Li},\ and\ \citenamefont
  {Carpentier}}]{Delplace2012}%
  \BibitemOpen
  \bibfield  {author} {\bibinfo {author} {\bibfnamefont {P.}~\bibnamefont
  {Delplace}}, \bibinfo {author} {\bibfnamefont {J.}~\bibnamefont {Li}}, \ and\
  \bibinfo {author} {\bibfnamefont {D.}~\bibnamefont {Carpentier}},\ }\href
  {\doibase 10.1209/0295-5075/97/67004} {\bibfield  {journal} {\bibinfo
  {journal} {{EPL} (Europhysics Letters)}\ }\textbf {\bibinfo {volume} {97}},\
  \bibinfo {pages} {67004} (\bibinfo {year} {2012})}\BibitemShut {NoStop}%
\bibitem [{\citenamefont {Xu}\ \emph {et~al.}(2015)\citenamefont {Xu},
  \citenamefont {Belopolski}, \citenamefont {Alidoust}, \citenamefont
  {Neupane}, \citenamefont {Bian}, \citenamefont {Zhang}, \citenamefont
  {Sankar}, \citenamefont {Chang}, \citenamefont {Yuan}, \citenamefont {Lee},
  \citenamefont {Huang}, \citenamefont {Zheng}, \citenamefont {Ma},
  \citenamefont {Sanchez}, \citenamefont {Wang}, \citenamefont {Bansil},
  \citenamefont {Chou}, \citenamefont {Shibayev}, \citenamefont {Lin},
  \citenamefont {Jia},\ and\ \citenamefont {Hasan}}]{Xu2015}%
  \BibitemOpen
  \bibfield  {author} {\bibinfo {author} {\bibfnamefont {S.-Y.}\ \bibnamefont
  {Xu}}, \bibinfo {author} {\bibfnamefont {I.}~\bibnamefont {Belopolski}},
  \bibinfo {author} {\bibfnamefont {N.}~\bibnamefont {Alidoust}}, \bibinfo
  {author} {\bibfnamefont {M.}~\bibnamefont {Neupane}}, \bibinfo {author}
  {\bibfnamefont {G.}~\bibnamefont {Bian}}, \bibinfo {author} {\bibfnamefont
  {C.}~\bibnamefont {Zhang}}, \bibinfo {author} {\bibfnamefont
  {R.}~\bibnamefont {Sankar}}, \bibinfo {author} {\bibfnamefont
  {G.}~\bibnamefont {Chang}}, \bibinfo {author} {\bibfnamefont
  {Z.}~\bibnamefont {Yuan}}, \bibinfo {author} {\bibfnamefont {C.-C.}\
  \bibnamefont {Lee}}, \bibinfo {author} {\bibfnamefont {S.-M.}\ \bibnamefont
  {Huang}}, \bibinfo {author} {\bibfnamefont {H.}~\bibnamefont {Zheng}},
  \bibinfo {author} {\bibfnamefont {J.}~\bibnamefont {Ma}}, \bibinfo {author}
  {\bibfnamefont {D.~S.}\ \bibnamefont {Sanchez}}, \bibinfo {author}
  {\bibfnamefont {B.}~\bibnamefont {Wang}}, \bibinfo {author} {\bibfnamefont
  {A.}~\bibnamefont {Bansil}}, \bibinfo {author} {\bibfnamefont
  {F.}~\bibnamefont {Chou}}, \bibinfo {author} {\bibfnamefont {P.~P.}\
  \bibnamefont {Shibayev}}, \bibinfo {author} {\bibfnamefont {H.}~\bibnamefont
  {Lin}}, \bibinfo {author} {\bibfnamefont {S.}~\bibnamefont {Jia}}, \ and\
  \bibinfo {author} {\bibfnamefont {M.~Z.}\ \bibnamefont {Hasan}},\ }\href
  {\doibase 10.1126/science.aaa9297} {\bibfield  {journal} {\bibinfo  {journal}
  {Science}\ }\textbf {\bibinfo {volume} {349}},\ \bibinfo {pages} {613}
  (\bibinfo {year} {2015})}\BibitemShut {NoStop}%
\bibitem [{\citenamefont {Fujita}\ \emph {et~al.}(1996)\citenamefont {Fujita},
  \citenamefont {Wakabayashi}, \citenamefont {Nakada},\ and\ \citenamefont
  {Kusakabe}}]{Fujita1996}%
  \BibitemOpen
  \bibfield  {author} {\bibinfo {author} {\bibfnamefont {M.}~\bibnamefont
  {Fujita}}, \bibinfo {author} {\bibfnamefont {K.}~\bibnamefont {Wakabayashi}},
  \bibinfo {author} {\bibfnamefont {K.}~\bibnamefont {Nakada}}, \ and\ \bibinfo
  {author} {\bibfnamefont {K.}~\bibnamefont {Kusakabe}},\ }\href {\doibase
  10.1143/JPSJ.65.1920} {\bibfield  {journal} {\bibinfo  {journal} {Journal of
  the Physical Society of Japan}\ }\textbf {\bibinfo {volume} {65}},\ \bibinfo
  {pages} {1920} (\bibinfo {year} {1996})}\BibitemShut {NoStop}%
\bibitem [{\citenamefont {Nakada}\ \emph {et~al.}(1996)\citenamefont {Nakada},
  \citenamefont {Fujita}, \citenamefont {Dresselhaus},\ and\ \citenamefont
  {Dresselhaus}}]{Nakada1996}%
  \BibitemOpen
  \bibfield  {author} {\bibinfo {author} {\bibfnamefont {K.}~\bibnamefont
  {Nakada}}, \bibinfo {author} {\bibfnamefont {M.}~\bibnamefont {Fujita}},
  \bibinfo {author} {\bibfnamefont {G.}~\bibnamefont {Dresselhaus}}, \ and\
  \bibinfo {author} {\bibfnamefont {M.~S.}\ \bibnamefont {Dresselhaus}},\
  }\href {\doibase 10.1103/PhysRevB.54.17954} {\bibfield  {journal} {\bibinfo
  {journal} {Phys. Rev. B}\ }\textbf {\bibinfo {volume} {54}},\ \bibinfo
  {pages} {17954} (\bibinfo {year} {1996})}\BibitemShut {NoStop}%
\bibitem [{\citenamefont {Palumbo}\ and\ \citenamefont
  {Meichanetzidis}(2015)}]{Palumbo2015}%
  \BibitemOpen
  \bibfield  {author} {\bibinfo {author} {\bibfnamefont {G.}~\bibnamefont
  {Palumbo}}\ and\ \bibinfo {author} {\bibfnamefont {K.}~\bibnamefont
  {Meichanetzidis}},\ }\href {\doibase 10.1103/PhysRevB.92.235106} {\bibfield
  {journal} {\bibinfo  {journal} {Phys. Rev. B}\ }\textbf {\bibinfo {volume}
  {92}},\ \bibinfo {pages} {235106} (\bibinfo {year} {2015})}\BibitemShut
  {NoStop}%
\bibitem [{\citenamefont {Zhou}\ \emph {et~al.}(2016)\citenamefont {Zhou},
  \citenamefont {Chen},\ and\ \citenamefont {Gong}}]{Jiangbin2016}%
  \BibitemOpen
  \bibfield  {author} {\bibinfo {author} {\bibfnamefont {L.}~\bibnamefont
  {Zhou}}, \bibinfo {author} {\bibfnamefont {C.}~\bibnamefont {Chen}}, \ and\
  \bibinfo {author} {\bibfnamefont {J.}~\bibnamefont {Gong}},\ }\href {\doibase
  10.1103/PhysRevB.94.075443} {\bibfield  {journal} {\bibinfo  {journal} {Phys.
  Rev. B}\ }\textbf {\bibinfo {volume} {94}},\ \bibinfo {pages} {075443}
  (\bibinfo {year} {2016})}\BibitemShut {NoStop}%
\bibitem [{\citenamefont {Ying}\ and\ \citenamefont
  {Kamenev}(2018)}]{Kamenev2018}%
  \BibitemOpen
  \bibfield  {author} {\bibinfo {author} {\bibfnamefont {X.}~\bibnamefont
  {Ying}}\ and\ \bibinfo {author} {\bibfnamefont {A.}~\bibnamefont {Kamenev}},\
  }\href {\doibase 10.1103/PhysRevLett.121.086810} {\bibfield  {journal}
  {\bibinfo  {journal} {Phys. Rev. Lett.}\ }\textbf {\bibinfo {volume} {121}},\
  \bibinfo {pages} {086810} (\bibinfo {year} {2018})}\BibitemShut {NoStop}%
\bibitem [{\citenamefont {Hatsugai}(1993)}]{Hatsugai1993}%
  \BibitemOpen
  \bibfield  {author} {\bibinfo {author} {\bibfnamefont {Y.}~\bibnamefont
  {Hatsugai}},\ }\href {\doibase 10.1103/PhysRevLett.71.3697} {\bibfield
  {journal} {\bibinfo  {journal} {Phys. Rev. Lett.}\ }\textbf {\bibinfo
  {volume} {71}},\ \bibinfo {pages} {3697} (\bibinfo {year}
  {1993})}\BibitemShut {NoStop}%
\bibitem [{\citenamefont {Graf}\ and\ \citenamefont {Porta}(2013)}]{Graf2013}%
  \BibitemOpen
  \bibfield  {author} {\bibinfo {author} {\bibfnamefont {G.~M.}\ \bibnamefont
  {Graf}}\ and\ \bibinfo {author} {\bibfnamefont {M.}~\bibnamefont {Porta}},\
  }\href {\doibase 10.1007/s00220-013-1819-6} {\bibfield  {journal} {\bibinfo
  {journal} {Communications in Mathematical Physics}\ }\textbf {\bibinfo
  {volume} {324}},\ \bibinfo {pages} {851} (\bibinfo {year}
  {2013})}\BibitemShut {NoStop}%
\bibitem [{\citenamefont {Kitagawa}\ \emph {et~al.}(2012)\citenamefont
  {Kitagawa}, \citenamefont {Broome}, \citenamefont {Fedrizzi}, \citenamefont
  {Rudner}, \citenamefont {Berg}, \citenamefont {Kassal}, \citenamefont
  {Aspuru-Guzik}, \citenamefont {Demler},\ and\ \citenamefont
  {White}}]{Kitagawa2012}%
  \BibitemOpen
  \bibfield  {author} {\bibinfo {author} {\bibfnamefont {T.}~\bibnamefont
  {Kitagawa}}, \bibinfo {author} {\bibfnamefont {M.~A.}\ \bibnamefont
  {Broome}}, \bibinfo {author} {\bibfnamefont {A.}~\bibnamefont {Fedrizzi}},
  \bibinfo {author} {\bibfnamefont {M.~S.}\ \bibnamefont {Rudner}}, \bibinfo
  {author} {\bibfnamefont {E.}~\bibnamefont {Berg}}, \bibinfo {author}
  {\bibfnamefont {I.}~\bibnamefont {Kassal}}, \bibinfo {author} {\bibfnamefont
  {A.}~\bibnamefont {Aspuru-Guzik}}, \bibinfo {author} {\bibfnamefont
  {E.}~\bibnamefont {Demler}}, \ and\ \bibinfo {author} {\bibfnamefont {A.~G.}\
  \bibnamefont {White}},\ }\href {\doibase 10.1038/ncomms1872} {\bibfield
  {journal} {\bibinfo  {journal} {Nat. Commun.}\ }\textbf {\bibinfo {volume}
  {3}},\ \bibinfo {pages} {882} (\bibinfo {year} {2012})}\BibitemShut {NoStop}%
\bibitem [{\citenamefont {Bellec}\ \emph {et~al.}(2017)\citenamefont {Bellec},
  \citenamefont {Michel}, \citenamefont {Zhang}, \citenamefont {Tzortzakis},\
  and\ \citenamefont {Delplace}}]{Bellec2017}%
  \BibitemOpen
  \bibfield  {author} {\bibinfo {author} {\bibfnamefont {M.}~\bibnamefont
  {Bellec}}, \bibinfo {author} {\bibfnamefont {C.}~\bibnamefont {Michel}},
  \bibinfo {author} {\bibfnamefont {H.}~\bibnamefont {Zhang}}, \bibinfo
  {author} {\bibfnamefont {S.}~\bibnamefont {Tzortzakis}}, \ and\ \bibinfo
  {author} {\bibfnamefont {P.}~\bibnamefont {Delplace}},\ }\href
  {https://iopscience.iop.org/article/10.1209/0295-5075/119/14003} {\bibfield
  {journal} {\bibinfo  {journal} {EPL (Europhysics Letters)}\ }\textbf
  {\bibinfo {volume} {119}},\ \bibinfo {pages} {14003} (\bibinfo {year}
  {2017})}\BibitemShut {NoStop}%
\bibitem [{\citenamefont {Wimmer}\ \emph {et~al.}(2017)\citenamefont {Wimmer},
  \citenamefont {Price}, \citenamefont {Carusotto},\ and\ \citenamefont
  {Peschel}}]{Wimmer2017}%
  \BibitemOpen
  \bibfield  {author} {\bibinfo {author} {\bibfnamefont {M.}~\bibnamefont
  {Wimmer}}, \bibinfo {author} {\bibfnamefont {H.~M.}\ \bibnamefont {Price}},
  \bibinfo {author} {\bibfnamefont {I.}~\bibnamefont {Carusotto}}, \ and\
  \bibinfo {author} {\bibfnamefont {U.}~\bibnamefont {Peschel}},\ }\href
  {https://www.nature.com/articles/nphys4050} {\bibfield  {journal} {\bibinfo
  {journal} {Nat. Phys.}\ }\textbf {\bibinfo {volume} {13}},\ \bibinfo {pages}
  {545} (\bibinfo {year} {2017})}\BibitemShut {NoStop}%
\bibitem [{\citenamefont {Liang}\ and\ \citenamefont
  {Chong}(2013)}]{Chong2013}%
  \BibitemOpen
  \bibfield  {author} {\bibinfo {author} {\bibfnamefont {G.~Q.}\ \bibnamefont
  {Liang}}\ and\ \bibinfo {author} {\bibfnamefont {Y.~D.}\ \bibnamefont
  {Chong}},\ }\href {\doibase 10.1103/PhysRevLett.110.203904} {\bibfield
  {journal} {\bibinfo  {journal} {Phys. Rev. Lett.}\ }\textbf {\bibinfo
  {volume} {110}},\ \bibinfo {pages} {203904} (\bibinfo {year}
  {2013})}\BibitemShut {NoStop}%
\bibitem [{\citenamefont {Pasek}\ and\ \citenamefont
  {Chong}(2014)}]{Chong2014}%
  \BibitemOpen
  \bibfield  {author} {\bibinfo {author} {\bibfnamefont {M.}~\bibnamefont
  {Pasek}}\ and\ \bibinfo {author} {\bibfnamefont {Y.~D.}\ \bibnamefont
  {Chong}},\ }\href {\doibase 10.1103/PhysRevB.89.075113} {\bibfield  {journal}
  {\bibinfo  {journal} {Phys. Rev. B}\ }\textbf {\bibinfo {volume} {89}},\
  \bibinfo {pages} {075113} (\bibinfo {year} {2014})}\BibitemShut {NoStop}%
\bibitem [{\citenamefont {Hu}\ \emph {et~al.}(2015)\citenamefont {Hu},
  \citenamefont {Pillay}, \citenamefont {Wu}, \citenamefont {Pasek},
  \citenamefont {Shum},\ and\ \citenamefont {Chong}}]{Chong2015X}%
  \BibitemOpen
  \bibfield  {author} {\bibinfo {author} {\bibfnamefont {W.}~\bibnamefont
  {Hu}}, \bibinfo {author} {\bibfnamefont {J.~C.}\ \bibnamefont {Pillay}},
  \bibinfo {author} {\bibfnamefont {K.}~\bibnamefont {Wu}}, \bibinfo {author}
  {\bibfnamefont {M.}~\bibnamefont {Pasek}}, \bibinfo {author} {\bibfnamefont
  {P.~P.}\ \bibnamefont {Shum}}, \ and\ \bibinfo {author} {\bibfnamefont
  {Y.~D.}\ \bibnamefont {Chong}},\ }\href {\doibase 10.1103/PhysRevX.5.011012}
  {\bibfield  {journal} {\bibinfo  {journal} {Phys. Rev. X}\ }\textbf {\bibinfo
  {volume} {5}},\ \bibinfo {pages} {011012} (\bibinfo {year}
  {2015})}\BibitemShut {NoStop}%
\bibitem [{\citenamefont {Tauber}\ and\ \citenamefont
  {Delplace}(2015)}]{Tauber2015}%
  \BibitemOpen
  \bibfield  {author} {\bibinfo {author} {\bibfnamefont {C.}~\bibnamefont
  {Tauber}}\ and\ \bibinfo {author} {\bibfnamefont {P.}~\bibnamefont
  {Delplace}},\ }\href@noop {} {\bibfield  {journal} {\bibinfo  {journal} {New
  J. Phys.}\ }\textbf {\bibinfo {volume} {17}},\ \bibinfo {pages} {115008}
  (\bibinfo {year} {2015})}\BibitemShut {NoStop}%
\bibitem [{\citenamefont {Wang}\ \emph {et~al.}(2016)\citenamefont {Wang},
  \citenamefont {Zhou},\ and\ \citenamefont {Chong}}]{Chong2016}%
  \BibitemOpen
  \bibfield  {author} {\bibinfo {author} {\bibfnamefont {H.}~\bibnamefont
  {Wang}}, \bibinfo {author} {\bibfnamefont {L.}~\bibnamefont {Zhou}}, \ and\
  \bibinfo {author} {\bibfnamefont {Y.~D.}\ \bibnamefont {Chong}},\ }\href
  {\doibase 10.1103/PhysRevB.93.144114} {\bibfield  {journal} {\bibinfo
  {journal} {Phys. Rev. B}\ }\textbf {\bibinfo {volume} {93}},\ \bibinfo
  {pages} {144114} (\bibinfo {year} {2016})}\BibitemShut {NoStop}%
\bibitem [{\citenamefont {Delplace}\ \emph
  {et~al.}(2017{\natexlab{a}})\citenamefont {Delplace}, \citenamefont
  {Fruchart},\ and\ \citenamefont {Tauber}}]{Delplace2017}%
  \BibitemOpen
  \bibfield  {author} {\bibinfo {author} {\bibfnamefont {P.}~\bibnamefont
  {Delplace}}, \bibinfo {author} {\bibfnamefont {M.}~\bibnamefont {Fruchart}},
  \ and\ \bibinfo {author} {\bibfnamefont {C.}~\bibnamefont {Tauber}},\ }\href
  {\doibase 10.1103/PhysRevB.95.205413} {\bibfield  {journal} {\bibinfo
  {journal} {Phys. Rev. B}\ }\textbf {\bibinfo {volume} {95}},\ \bibinfo
  {pages} {205413} (\bibinfo {year} {2017}{\natexlab{a}})}\BibitemShut
  {NoStop}%
\bibitem [{\citenamefont {Delplace}(2019)}]{Delplace2019}%
  \BibitemOpen
  \bibfield  {author} {\bibinfo {author} {\bibfnamefont {P.}~\bibnamefont
  {Delplace}},\ }\href {https://arxiv.org/abs/1905.11194} {\  (\bibinfo {year}
  {2019})},\ \Eprint {http://arxiv.org/abs/1905.11194} {arXiv:1905.11194}
  \BibitemShut {NoStop}%
\bibitem [{\citenamefont {Upreti}\ \emph {et~al.}(2019)\citenamefont {Upreti},
  \citenamefont {Evain}, \citenamefont {Randoux}, \citenamefont {Suret},
  \citenamefont {Amo},\ and\ \citenamefont {Delplace}}]{Upreti2019}%
  \BibitemOpen
  \bibfield  {author} {\bibinfo {author} {\bibfnamefont {L.~K.}\ \bibnamefont
  {Upreti}}, \bibinfo {author} {\bibfnamefont {C.}~\bibnamefont {Evain}},
  \bibinfo {author} {\bibfnamefont {S.}~\bibnamefont {Randoux}}, \bibinfo
  {author} {\bibfnamefont {P.}~\bibnamefont {Suret}}, \bibinfo {author}
  {\bibfnamefont {A.}~\bibnamefont {Amo}}, \ and\ \bibinfo {author}
  {\bibfnamefont {P.}~\bibnamefont {Delplace}},\ }\href@noop {} {\enquote
  {\bibinfo {title} {Floquet winding metals},}\ } (\bibinfo {year} {2019}),\
  \Eprint {http://arxiv.org/abs/1907.09914} {arXiv:1907.09914
  [cond-mat.mes-hall]} \BibitemShut {NoStop}%
\bibitem [{\citenamefont {Potter}\ \emph {et~al.}(2020)\citenamefont {Potter},
  \citenamefont {Chalker},\ and\ \citenamefont {Gurarie}}]{Potter2020}%
  \BibitemOpen
  \bibfield  {author} {\bibinfo {author} {\bibfnamefont {A.~C.}\ \bibnamefont
  {Potter}}, \bibinfo {author} {\bibfnamefont {J.}~\bibnamefont {Chalker}}, \
  and\ \bibinfo {author} {\bibfnamefont {V.}~\bibnamefont {Gurarie}},\ }\href
  {https://arxiv.org/abs/2002.04058} {\bibfield  {journal} {\bibinfo  {journal}
  {arXiv preprint arXiv:2002.04058}\ } (\bibinfo {year} {2020})}\BibitemShut
  {NoStop}%
\bibitem [{\citenamefont {Rudner}\ \emph {et~al.}(2013)\citenamefont {Rudner},
  \citenamefont {Lindner}, \citenamefont {Berg},\ and\ \citenamefont
  {Levin}}]{Rudner2013}%
  \BibitemOpen
  \bibfield  {author} {\bibinfo {author} {\bibfnamefont {M.~S.}\ \bibnamefont
  {Rudner}}, \bibinfo {author} {\bibfnamefont {N.~H.}\ \bibnamefont {Lindner}},
  \bibinfo {author} {\bibfnamefont {E.}~\bibnamefont {Berg}}, \ and\ \bibinfo
  {author} {\bibfnamefont {M.}~\bibnamefont {Levin}},\ }\href {\doibase
  10.1103/PhysRevX.3.031005} {\bibfield  {journal} {\bibinfo  {journal} {Phys.
  Rev. X}\ }\textbf {\bibinfo {volume} {3}},\ \bibinfo {pages} {031005}
  (\bibinfo {year} {2013})}\BibitemShut {NoStop}%
\bibitem [{\citenamefont {Upreti}\ \emph {et~al.}(2020)\citenamefont {Upreti},
  \citenamefont {Evain}, \citenamefont {Randoux}, \citenamefont {Suret},
  \citenamefont {Amo},\ and\ \citenamefont {Delplace}}]{Upreti2020}%
  \BibitemOpen
  \bibfield  {author} {\bibinfo {author} {\bibfnamefont {L.~K.}\ \bibnamefont
  {Upreti}}, \bibinfo {author} {\bibfnamefont {C.}~\bibnamefont {Evain}},
  \bibinfo {author} {\bibfnamefont {S.}~\bibnamefont {Randoux}}, \bibinfo
  {author} {\bibfnamefont {P.}~\bibnamefont {Suret}}, \bibinfo {author}
  {\bibfnamefont {A.}~\bibnamefont {Amo}}, \ and\ \bibinfo {author}
  {\bibfnamefont {P.}~\bibnamefont {Delplace}},\ }\href@noop {} {\enquote
  {\bibinfo {title} {Topological swing in bloch oscillations},}\ } (\bibinfo
  {year} {2020}),\ \Eprint {http://arxiv.org/abs/2004.14261} {arXiv:2004.14261
  [cond-mat.mes-hall]} \BibitemShut {NoStop}%
\bibitem [{\citenamefont {Banerjee}\ \emph {et~al.}(2009)\citenamefont
  {Banerjee}, \citenamefont {Singh}, \citenamefont {Pardo},\ and\ \citenamefont
  {Pickett}}]{Pickett}%
  \BibitemOpen
  \bibfield  {author} {\bibinfo {author} {\bibfnamefont {S.}~\bibnamefont
  {Banerjee}}, \bibinfo {author} {\bibfnamefont {R.~R.~P.}\ \bibnamefont
  {Singh}}, \bibinfo {author} {\bibfnamefont {V.}~\bibnamefont {Pardo}}, \ and\
  \bibinfo {author} {\bibfnamefont {W.~E.}\ \bibnamefont {Pickett}},\ }\href
  {\doibase 10.1103/PhysRevLett.103.016402} {\bibfield  {journal} {\bibinfo
  {journal} {Phys. Rev. Lett.}\ }\textbf {\bibinfo {volume} {103}},\ \bibinfo
  {pages} {016402} (\bibinfo {year} {2009})}\BibitemShut {NoStop}%
\bibitem [{\citenamefont {Montambaux}\ \emph {et~al.}(2009)\citenamefont
  {Montambaux}, \citenamefont {Pi\'echon}, \citenamefont {Fuchs},\ and\
  \citenamefont {Goerbig}}]{Gilles}%
  \BibitemOpen
  \bibfield  {author} {\bibinfo {author} {\bibfnamefont {G.}~\bibnamefont
  {Montambaux}}, \bibinfo {author} {\bibfnamefont {F.}~\bibnamefont
  {Pi\'echon}}, \bibinfo {author} {\bibfnamefont {J.-N.}\ \bibnamefont
  {Fuchs}}, \ and\ \bibinfo {author} {\bibfnamefont {M.~O.}\ \bibnamefont
  {Goerbig}},\ }\href {\doibase 10.1103/PhysRevB.80.153412} {\bibfield
  {journal} {\bibinfo  {journal} {Phys. Rev. B}\ }\textbf {\bibinfo {volume}
  {80}},\ \bibinfo {pages} {153412} (\bibinfo {year} {2009})}\BibitemShut
  {NoStop}%
\bibitem [{\citenamefont {Huang}\ \emph {et~al.}(2015)\citenamefont {Huang},
  \citenamefont {Liu}, \citenamefont {Zhang}, \citenamefont {Duan},\ and\
  \citenamefont {Vanderbilt}}]{Vanderbilt}%
  \BibitemOpen
  \bibfield  {author} {\bibinfo {author} {\bibfnamefont {H.}~\bibnamefont
  {Huang}}, \bibinfo {author} {\bibfnamefont {Z.}~\bibnamefont {Liu}}, \bibinfo
  {author} {\bibfnamefont {H.}~\bibnamefont {Zhang}}, \bibinfo {author}
  {\bibfnamefont {W.}~\bibnamefont {Duan}}, \ and\ \bibinfo {author}
  {\bibfnamefont {D.}~\bibnamefont {Vanderbilt}},\ }\href {\doibase
  10.1103/PhysRevB.92.161115} {\bibfield  {journal} {\bibinfo  {journal} {Phys.
  Rev. B}\ }\textbf {\bibinfo {volume} {92}},\ \bibinfo {pages} {161115}
  (\bibinfo {year} {2015})}\BibitemShut {NoStop}%
\bibitem [{\citenamefont {Banerjee}(2015)}]{Banerjee2015}%
  \BibitemOpen
  \bibfield  {author} {\bibinfo {author} {\bibfnamefont {S.}~\bibnamefont
  {Banerjee}},\ }\href {https://arxiv.org/abs/1508.05145} {\enquote {\bibinfo
  {title} {Anderson localizaion for semi-dirac semi-weyl semi-metal},}\ }
  (\bibinfo {year} {2015}),\ \Eprint {http://arxiv.org/abs/1508.05145}
  {arXiv:1508.05145 [cond-mat.str-el]} \BibitemShut {NoStop}%
\bibitem [{\citenamefont {Zhong}\ \emph {et~al.}(2017)\citenamefont {Zhong},
  \citenamefont {Chen}, \citenamefont {Xie}, \citenamefont {Sun},\ and\
  \citenamefont {Zhang}}]{Zhong2017}%
  \BibitemOpen
  \bibfield  {author} {\bibinfo {author} {\bibfnamefont {C.}~\bibnamefont
  {Zhong}}, \bibinfo {author} {\bibfnamefont {Y.}~\bibnamefont {Chen}},
  \bibinfo {author} {\bibfnamefont {Y.}~\bibnamefont {Xie}}, \bibinfo {author}
  {\bibfnamefont {Y.-Y.}\ \bibnamefont {Sun}}, \ and\ \bibinfo {author}
  {\bibfnamefont {S.}~\bibnamefont {Zhang}},\ }\href {\doibase
  10.1039/C6CP08439G} {\bibfield  {journal} {\bibinfo  {journal} {Phys. Chem.
  Chem. Phys.}\ }\textbf {\bibinfo {volume} {19}},\ \bibinfo {pages} {3820}
  (\bibinfo {year} {2017})}\BibitemShut {NoStop}%
\bibitem [{\citenamefont {Mawrie}\ and\ \citenamefont
  {Muralidharan}(2019)}]{Mawrie2019}%
  \BibitemOpen
  \bibfield  {author} {\bibinfo {author} {\bibfnamefont {A.}~\bibnamefont
  {Mawrie}}\ and\ \bibinfo {author} {\bibfnamefont {B.}~\bibnamefont
  {Muralidharan}},\ }\href {\doibase 10.1103/PhysRevB.100.081403} {\bibfield
  {journal} {\bibinfo  {journal} {Phys. Rev. B}\ }\textbf {\bibinfo {volume}
  {100}},\ \bibinfo {pages} {081403} (\bibinfo {year} {2019})}\BibitemShut
  {NoStop}%
\bibitem [{\citenamefont {Nakahara}(2003)}]{Nakahara2003}%
  \BibitemOpen
  \bibfield  {author} {\bibinfo {author} {\bibfnamefont {M.}~\bibnamefont
  {Nakahara}},\ }\href
  {https://www.crcpress.com/Geometry-Topology-and-Physics-Second-Edition/Nakahara/p/book/9780750306065}
  {\emph {\bibinfo {title} {Geometry, topology and physics}}}\ (\bibinfo
  {publisher} {CRC Press},\ \bibinfo {year} {2003})\BibitemShut {NoStop}%
\bibitem [{\citenamefont {Volovik}(2009)}]{Volovik2009}%
  \BibitemOpen
  \bibfield  {author} {\bibinfo {author} {\bibfnamefont {G.~E.}\ \bibnamefont
  {Volovik}},\ }\href {\doibase 10.1093/acprof:oso/9780199564842.001.0001}
  {\emph {\bibinfo {title} {Technology}}}\ (\bibinfo  {publisher} {Oxford
  University Press},\ \bibinfo {year} {2009})\BibitemShut {NoStop}%
\bibitem [{\citenamefont {Delplace}\ \emph
  {et~al.}(2017{\natexlab{b}})\citenamefont {Delplace}, \citenamefont
  {Marston},\ and\ \citenamefont {Venaille}}]{Delplace2017a}%
  \BibitemOpen
  \bibfield  {author} {\bibinfo {author} {\bibfnamefont {P.}~\bibnamefont
  {Delplace}}, \bibinfo {author} {\bibfnamefont {J.~B.}\ \bibnamefont
  {Marston}}, \ and\ \bibinfo {author} {\bibfnamefont {A.}~\bibnamefont
  {Venaille}},\ }\href {\doibase 10.1126/science.aan8819} {\bibfield  {journal}
  {\bibinfo  {journal} {Science}\ }\textbf {\bibinfo {volume} {358}},\ \bibinfo
  {pages} {1075} (\bibinfo {year} {2017}{\natexlab{b}})}\BibitemShut {NoStop}%
\bibitem [{\citenamefont {Faure}(2019)}]{Faure2019}%
  \BibitemOpen
  \bibfield  {author} {\bibinfo {author} {\bibfnamefont {F.}~\bibnamefont
  {Faure}},\ }\href {https://arxiv.org/abs/1901.10592} {\bibfield  {journal}
  {\bibinfo  {journal} {arXiv preprint arXiv:1901.10592}\ } (\bibinfo {year}
  {2019})}\BibitemShut {NoStop}%
\bibitem [{\citenamefont {Perrot}\ \emph {et~al.}(2019)\citenamefont {Perrot},
  \citenamefont {Delplace},\ and\ \citenamefont {Venaille}}]{Perrot2019}%
  \BibitemOpen
  \bibfield  {author} {\bibinfo {author} {\bibfnamefont {M.}~\bibnamefont
  {Perrot}}, \bibinfo {author} {\bibfnamefont {P.}~\bibnamefont {Delplace}}, \
  and\ \bibinfo {author} {\bibfnamefont {A.}~\bibnamefont {Venaille}},\ }\href
  {\doibase 10.1038/s41567-019-0561-1} {\bibfield  {journal} {\bibinfo
  {journal} {Nature Physics}\ }\textbf {\bibinfo {volume} {15}},\ \bibinfo
  {pages} {781} (\bibinfo {year} {2019})}\BibitemShut {NoStop}%
\bibitem [{\citenamefont {Marciani}\ and\ \citenamefont
  {Delplace}(2019)}]{Marciani2019}%
  \BibitemOpen
  \bibfield  {author} {\bibinfo {author} {\bibfnamefont {M.}~\bibnamefont
  {Marciani}}\ and\ \bibinfo {author} {\bibfnamefont {P.}~\bibnamefont
  {Delplace}},\ }\href {https://arxiv.org/abs/1906.09057} {\bibfield  {journal}
  {\bibinfo  {journal} {arXiv preprint arXiv:1906.09057}\ } (\bibinfo {year}
  {2019})}\BibitemShut {NoStop}%
\bibitem [{\citenamefont {Volovik}(1988)}]{Volovik1988}%
  \BibitemOpen
  \bibfield  {author} {\bibinfo {author} {\bibfnamefont {G.}~\bibnamefont
  {Volovik}},\ }\href
  {http://www.jetp.ac.ru/cgi-bin/e/index/e/67/9/p1804?a=list} {\bibfield
  {journal} {\bibinfo  {journal} {Zhurnal Ehksperimental'noj i Teoreticheskoj
  Fiziki}\ }\textbf {\bibinfo {volume} {20}} (\bibinfo {year}
  {1988})}\BibitemShut {NoStop}%
\bibitem [{\citenamefont {Silveirinha}(2015)}]{Silveirinha2015}%
  \BibitemOpen
  \bibfield  {author} {\bibinfo {author} {\bibfnamefont {M.~G.}\ \bibnamefont
  {Silveirinha}},\ }\href {\doibase 10.1103/PhysRevB.92.125153} {\bibfield
  {journal} {\bibinfo  {journal} {Phys. Rev. B}\ }\textbf {\bibinfo {volume}
  {92}},\ \bibinfo {pages} {125153} (\bibinfo {year} {2015})}\BibitemShut
  {NoStop}%
\bibitem [{\citenamefont {Tauber}\ \emph {et~al.}(2019)\citenamefont {Tauber},
  \citenamefont {Delplace},\ and\ \citenamefont {Venaille}}]{Tauber2019}%
  \BibitemOpen
  \bibfield  {author} {\bibinfo {author} {\bibfnamefont {C.}~\bibnamefont
  {Tauber}}, \bibinfo {author} {\bibfnamefont {P.}~\bibnamefont {Delplace}}, \
  and\ \bibinfo {author} {\bibfnamefont {A.}~\bibnamefont {Venaille}},\ }\href
  {\doibase 10.1017/jfm.2019.233} {\bibfield  {journal} {\bibinfo  {journal}
  {Journal of Fluid Mechanics}\ }\textbf {\bibinfo {volume} {868}},\ \bibinfo
  {pages} {R2} (\bibinfo {year} {2019})}\BibitemShut {NoStop}%
\bibitem [{\citenamefont {Souslov}\ \emph {et~al.}(2019)\citenamefont
  {Souslov}, \citenamefont {Dasbiswas}, \citenamefont {Fruchart}, \citenamefont
  {Vaikuntanathan},\ and\ \citenamefont {Vitelli}}]{Souslov2019}%
  \BibitemOpen
  \bibfield  {author} {\bibinfo {author} {\bibfnamefont {A.}~\bibnamefont
  {Souslov}}, \bibinfo {author} {\bibfnamefont {K.}~\bibnamefont {Dasbiswas}},
  \bibinfo {author} {\bibfnamefont {M.}~\bibnamefont {Fruchart}}, \bibinfo
  {author} {\bibfnamefont {S.}~\bibnamefont {Vaikuntanathan}}, \ and\ \bibinfo
  {author} {\bibfnamefont {V.}~\bibnamefont {Vitelli}},\ }\href {\doibase
  10.1103/PhysRevLett.122.128001} {\bibfield  {journal} {\bibinfo  {journal}
  {Phys. Rev. Lett.}\ }\textbf {\bibinfo {volume} {122}},\ \bibinfo {pages}
  {128001} (\bibinfo {year} {2019})}\BibitemShut {NoStop}%
\bibitem [{\citenamefont {Tauber}\ \emph {et~al.}(2020)\citenamefont {Tauber},
  \citenamefont {Delplace},\ and\ \citenamefont {Venaille}}]{Tauber2020}%
  \BibitemOpen
  \bibfield  {author} {\bibinfo {author} {\bibfnamefont {C.}~\bibnamefont
  {Tauber}}, \bibinfo {author} {\bibfnamefont {P.}~\bibnamefont {Delplace}}, \
  and\ \bibinfo {author} {\bibfnamefont {A.}~\bibnamefont {Venaille}},\ }\href
  {\doibase 10.1103/PhysRevResearch.2.013147} {\bibfield  {journal} {\bibinfo
  {journal} {Phys. Rev. Research}\ }\textbf {\bibinfo {volume} {2}},\ \bibinfo
  {pages} {013147} (\bibinfo {year} {2020})}\BibitemShut {NoStop}%
\bibitem [{\citenamefont {Wimmer}\ \emph {et~al.}(2013)\citenamefont {Wimmer},
  \citenamefont {Regensburger}, \citenamefont {Bersch}, \citenamefont {Miri},
  \citenamefont {Batz}, \citenamefont {Onishchukov}, \citenamefont
  {Christodoulides},\ and\ \citenamefont {Peschel}}]{Wimmer2013}%
  \BibitemOpen
  \bibfield  {author} {\bibinfo {author} {\bibfnamefont {M.}~\bibnamefont
  {Wimmer}}, \bibinfo {author} {\bibfnamefont {A.}~\bibnamefont
  {Regensburger}}, \bibinfo {author} {\bibfnamefont {C.}~\bibnamefont
  {Bersch}}, \bibinfo {author} {\bibfnamefont {M.-A.}\ \bibnamefont {Miri}},
  \bibinfo {author} {\bibfnamefont {S.}~\bibnamefont {Batz}}, \bibinfo {author}
  {\bibfnamefont {G.}~\bibnamefont {Onishchukov}}, \bibinfo {author}
  {\bibfnamefont {D.~N.}\ \bibnamefont {Christodoulides}}, \ and\ \bibinfo
  {author} {\bibfnamefont {U.}~\bibnamefont {Peschel}},\ }\href
  {http://dx.doi.org/10.1038/nphys2777} {\bibfield  {journal} {\bibinfo
  {journal} {Nat. Phys.}\ }\textbf {\bibinfo {volume} {9}},\ \bibinfo {pages}
  {780} (\bibinfo {year} {2013})}\BibitemShut {NoStop}%
\bibitem [{\citenamefont {Regensburger}\ \emph {et~al.}(2011)\citenamefont
  {Regensburger}, \citenamefont {Bersch}, \citenamefont {Hinrichs},
  \citenamefont {Onishchukov}, \citenamefont {Schreiber}, \citenamefont
  {Silberhorn},\ and\ \citenamefont {Peschel}}]{AloisPRL2011}%
  \BibitemOpen
  \bibfield  {author} {\bibinfo {author} {\bibfnamefont {A.}~\bibnamefont
  {Regensburger}}, \bibinfo {author} {\bibfnamefont {C.}~\bibnamefont
  {Bersch}}, \bibinfo {author} {\bibfnamefont {B.}~\bibnamefont {Hinrichs}},
  \bibinfo {author} {\bibfnamefont {G.}~\bibnamefont {Onishchukov}}, \bibinfo
  {author} {\bibfnamefont {A.}~\bibnamefont {Schreiber}}, \bibinfo {author}
  {\bibfnamefont {C.}~\bibnamefont {Silberhorn}}, \ and\ \bibinfo {author}
  {\bibfnamefont {U.}~\bibnamefont {Peschel}},\ }\href {\doibase
  10.1103/PhysRevLett.107.233902} {\bibfield  {journal} {\bibinfo  {journal}
  {Phys. Rev. Lett.}\ }\textbf {\bibinfo {volume} {107}},\ \bibinfo {pages}
  {233902} (\bibinfo {year} {2011})}\BibitemShut {NoStop}%
\bibitem [{\citenamefont {Miri}\ \emph {et~al.}(2012)\citenamefont {Miri},
  \citenamefont {Regensburger}, \citenamefont {Peschel},\ and\ \citenamefont
  {Christodoulides}}]{Miri2012}%
  \BibitemOpen
  \bibfield  {author} {\bibinfo {author} {\bibfnamefont {M.-A.}\ \bibnamefont
  {Miri}}, \bibinfo {author} {\bibfnamefont {A.}~\bibnamefont {Regensburger}},
  \bibinfo {author} {\bibfnamefont {U.}~\bibnamefont {Peschel}}, \ and\
  \bibinfo {author} {\bibfnamefont {D.~N.}\ \bibnamefont {Christodoulides}},\
  }\href {\doibase 10.1103/PhysRevA.86.023807} {\bibfield  {journal} {\bibinfo
  {journal} {Phys. Rev. A}\ }\textbf {\bibinfo {volume} {86}},\ \bibinfo
  {pages} {023807} (\bibinfo {year} {2012})}\BibitemShut {NoStop}%
\bibitem [{\citenamefont {Regensburger}\ \emph {et~al.}(2012)\citenamefont
  {Regensburger}, \citenamefont {Bersch}, \citenamefont {Miri}, \citenamefont
  {Onishchukov}, \citenamefont {Christodoulides},\ and\ \citenamefont
  {Peschel}}]{Regensburger2012}%
  \BibitemOpen
  \bibfield  {author} {\bibinfo {author} {\bibfnamefont {A.}~\bibnamefont
  {Regensburger}}, \bibinfo {author} {\bibfnamefont {C.}~\bibnamefont
  {Bersch}}, \bibinfo {author} {\bibfnamefont {M.-A.}\ \bibnamefont {Miri}},
  \bibinfo {author} {\bibfnamefont {G.}~\bibnamefont {Onishchukov}}, \bibinfo
  {author} {\bibfnamefont {D.~N.}\ \bibnamefont {Christodoulides}}, \ and\
  \bibinfo {author} {\bibfnamefont {U.}~\bibnamefont {Peschel}},\ }\href
  {\doibase 10.1038/nature11298} {\bibfield  {journal} {\bibinfo  {journal}
  {Nature}\ }\textbf {\bibinfo {volume} {488}},\ \bibinfo {pages} {167}
  (\bibinfo {year} {2012})}\BibitemShut {NoStop}%
\bibitem [{\citenamefont {Weidemann}\ \emph {et~al.}(2020)\citenamefont
  {Weidemann}, \citenamefont {Kremer}, \citenamefont {Helbig}, \citenamefont
  {Hofmann}, \citenamefont {Stegmaier}, \citenamefont {Greiter}, \citenamefont
  {Thomale},\ and\ \citenamefont {Szameit}}]{Weidemann2020}%
  \BibitemOpen
  \bibfield  {author} {\bibinfo {author} {\bibfnamefont {S.}~\bibnamefont
  {Weidemann}}, \bibinfo {author} {\bibfnamefont {M.}~\bibnamefont {Kremer}},
  \bibinfo {author} {\bibfnamefont {T.}~\bibnamefont {Helbig}}, \bibinfo
  {author} {\bibfnamefont {T.}~\bibnamefont {Hofmann}}, \bibinfo {author}
  {\bibfnamefont {A.}~\bibnamefont {Stegmaier}}, \bibinfo {author}
  {\bibfnamefont {M.}~\bibnamefont {Greiter}}, \bibinfo {author} {\bibfnamefont
  {R.}~\bibnamefont {Thomale}}, \ and\ \bibinfo {author} {\bibfnamefont
  {A.}~\bibnamefont {Szameit}},\ }\href {\doibase 10.1126/science.aaz8727}
  {\bibfield  {journal} {\bibinfo  {journal} {Science}\ }\textbf {\bibinfo
  {volume} {368}},\ \bibinfo {pages} {311} (\bibinfo {year}
  {2020})}\BibitemShut {NoStop}%
\bibitem [{\citenamefont {Szameit}\ and\ \citenamefont
  {Nolte}(2010)}]{Szameit2010}%
  \BibitemOpen
  \bibfield  {author} {\bibinfo {author} {\bibfnamefont {A.}~\bibnamefont
  {Szameit}}\ and\ \bibinfo {author} {\bibfnamefont {S.}~\bibnamefont
  {Nolte}},\ }\href {\doibase 10.1088/0953-4075/43/16/163001} {\bibfield
  {journal} {\bibinfo  {journal} {Journal of Physics B: Atomic, Molecular and
  Optical Physics}\ }\textbf {\bibinfo {volume} {43}} (\bibinfo {year}
  {2010}),\ 10.1088/0953-4075/43/16/163001}\BibitemShut {NoStop}%
\bibitem [{\citenamefont {Rechtsman}\ \emph {et~al.}(2013)\citenamefont
  {Rechtsman}, \citenamefont {Zeuner}, \citenamefont {Plotnik}, \citenamefont
  {Lumer}, \citenamefont {Podolsky}, \citenamefont {Dreisow}, \citenamefont
  {Nolte}, \citenamefont {Segev},\ and\ \citenamefont
  {Szameit}}]{Rechtsman2013}%
  \BibitemOpen
  \bibfield  {author} {\bibinfo {author} {\bibfnamefont {M.~C.}\ \bibnamefont
  {Rechtsman}}, \bibinfo {author} {\bibfnamefont {J.~M.}\ \bibnamefont
  {Zeuner}}, \bibinfo {author} {\bibfnamefont {Y.}~\bibnamefont {Plotnik}},
  \bibinfo {author} {\bibfnamefont {Y.}~\bibnamefont {Lumer}}, \bibinfo
  {author} {\bibfnamefont {D.}~\bibnamefont {Podolsky}}, \bibinfo {author}
  {\bibfnamefont {F.}~\bibnamefont {Dreisow}}, \bibinfo {author} {\bibfnamefont
  {S.}~\bibnamefont {Nolte}}, \bibinfo {author} {\bibfnamefont
  {M.}~\bibnamefont {Segev}}, \ and\ \bibinfo {author} {\bibfnamefont
  {A.}~\bibnamefont {Szameit}},\ }\href {\doibase 10.1038/nature12066}
  {\bibfield  {journal} {\bibinfo  {journal} {Nature}\ }\textbf {\bibinfo
  {volume} {496}},\ \bibinfo {pages} {196} (\bibinfo {year}
  {2013})}\BibitemShut {NoStop}%
\end{thebibliography}%

\end{document}